\documentclass[12pt,doublespace,onecolumn]{IEEEtran}
\usepackage[dvips]{epsfig}
\usepackage[dvips]{graphicx}

\usepackage{amsmath,amsfonts,bm,amssymb,psfrag,ifthen,color,subfigure}
\usepackage{algpseudocode}
\usepackage{algorithmicx}
\usepackage{setspace}
\usepackage{cite}
\usepackage{longtable}
\usepackage{stfloats}  
\usepackage{graphics,booktabs,color,subfigure}

\newcommand\xb{\ensuremath{{\bf x}}}

\newcommand\yb{\ensuremath{{\bm y}}}

\newcommand\rb{\ensuremath{{\bf r}}}

\newcommand\Cplx{\ensuremath{{\mathbb{C}}}}

\newcommand{\cL}{\mathcal{L}}

\newcommand{\cQ}{\mathcal{Q}}

\newcommand{\cN}{\mathcal{N}}

\newcommand{\cC}{\mathcal{C}}

\newcommand{\bfalpha}{{\mbox{\boldmath $\alpha$}}}

\newcommand{\bfdelta}{{\mbox{\boldmath $\delta$}}}
\newcommand{\bftheta}{{\mbox{\boldmath $\theta$}}}
\newcommand{\bfepsilon}{{\mbox{\boldmath $\epsilon$}}}
\newcommand{\bfkappa}{{\mbox{\boldmath $\kappa$}}}

\newtheorem{lemma}{Lemma}

\newtheorem{Theorem}{Theorem}

\newtheorem{Rmk}{Remark}
\def\thesection{\arabic{section}}

\title{Min Flow Rate Maximization for Software Defined Radio Access Networks}
\author{Wei-Cheng Liao, Mingyi Hong, Hamid Farmanbar, Xu Li, Zhi-Quan Luo, and Hang Zhang
\thanks{This work is supported in part by NSF, grant number CCF-1216858, and in part by a research gift from Huawei Technologies Inc.}\thanks{The conference version of this manuscript has been submitted to ICASSP 2014. \cite{Liao14SDN}}
\thanks{W.-C. Liao, M. Hong and Z.-Q. Luo are with the Department of Electrical and Computer Engineering, University of Minnesota, Minneapolis, MN 55455, USA}
\thanks{H. Farmanbar, X. Li, and H. Zhang are with the Ottawa R\&D Centre, Huawei Technologies Canada, Ottawa, ON, Canada}}

\begin{document}
\maketitle
\vspace{-0.6cm}
\begin{abstract}
We consider a heterogeneous network (HetNet) of base stations (BSs)
connected via a backhaul network of routers and wired/wireless links
with limited capacity. The optimal provision of such networks
requires proper resource allocation across the radio access links in
conjunction with appropriate traffic engineering within the backhaul
network. In this paper we propose an efficient algorithm for joint
resource allocation across the wireless links and the flow control
within the backhaul network. The proposed algorithm, which maximizes
the minimum rate among all the users and/or flows, is based on a
decomposition approach that leverages both the Alternating Direction
Method of Multipliers (ADMM) and the weighted-MMSE (WMMSE)
algorithm. We show that this algorithm is easily parallelizable and
converges globally to a stationary solution of the joint
optimization problem. The proposed algorithm can also be extended to
deal with per-flow quality of service constraint, or to networks
with multi-antenna nodes.

\vspace{-0.1cm}
\end{abstract}
\vspace{-0.4cm}
\section{Introduction}
\label{sec:intro}

With the advent of cloud computing technologies and the mass
deployment of low power base stations (BSs), the cellular radio
access networks (RAN) has undergone a major structural change.
The traditional high powered single-hop access mode between a serving BS
and its users is being replaced by a mesh network consisting of a
large number of wireless access points connected by either wireline
or wireless backhaul links as well as network routers \cite{Andrews13}.
New concepts such as heterogeneous network (HetNet) or software defined air interface  that capture
these changes have been proposed and studied recently
(see \cite{Gesbert10, HuaweiSDN} and references therein).
Such cloud-based, software defined RAN (SD-RAN) architecture has been envisioned as a future 5G standard, and
is expected to achieve 1000x performance improvement over the current
4G technology within the next ten years \cite{HuaweiSDN}.

The success of the software defined radio access networks will
depend critically on our ability to jointly provision the backhaul traffic and mitigate interference in the air interface.
In recent years, interference management has been a
major focus of the wireless communication research \cite{hong12survey,Bjornson13}. For instance, various downlink interference
management techniques have been developed under the assumption that
the wireless user data can be routed to the transmitting BSs without any cost to the backhaul network. Unfortunately, such idealized assumption is only reasonable for traditional networks with a small number of networked BSs for which traffic engineering is straightforward. In the next generation RAN, there will be a large number of BSs, many of which may be connected to the core network without carrier-grade backhaul, e.g., WIFI access points with digital subscriber line (DSL) connections. The increased heterogeneity, network size and backhaul constraints make
interference management for future cloud based RANs a challenging task.

As a multi-commodity flow problem, backhaul traffic engineering involves multi-hop routing
from the source nodes (e.g., the cloud centers with backhaul
connection) to the destination nodes (e.g., the users requesting
content). The resulting optimal solution must guarantee the requested quality of service (QoS) for each
end-to-end flow (or commodities in the terminology of traffic
engineering) while satisfying the capacity constraints for all the wireless and/or wired links used by the flows.
Compared to the traditional multi-commodity routing in
wireline networks \cite{Bertsekas87,Bertsekas88}, traffic engineering in the wireless
setting is much more challenging due to several reasons. First, the link capacity between two nearby nodes
is a nonconvex function of the transmit power budget, channel strength, as well as the underlying
physical layer coding/decoding techniques used. Second, the amount of
traffic that can be carried on neighboring links is interdependent
due to the multiuser interference caused by nearby nodes. Third,
multiple parallel channels between two nodes may be available for
transmission. To respond to these new challenges, novel RAN management
methods must be developed for joint wireless resource allocation in the air interface and
traffic engineering within the multi-hop backhaul network. These
methods together will be a central component of the newly proposed software defined networking (SDN)
concept \cite{HuaweiSDN,SDN}, which advocates centralized network
provisioning for cloud based radio access networks.

The impact of the finite bandwidth of backhaul networks on wireless
resource allocation has been studied recently in the context of
joint processing between BSs, e.g.,
\cite{Zakhour11,Chowdhery11,Mehryar12,Park13}. However, these works
do not consider multi-hop routing between the
source and the destination nodes. The joint optimization of the backhaul flow
routing and the power allocation for
wireless network has also been considered in the framework of
cross-layer network utility maximization (NUM) problem, see e.g.
\cite{Xiao04,Neely05,Ribeiro10,Shao11} and some tutorial papers
\cite{Georgiadis06,Chiang07,Shroff06}. However, since the
capacity of wireless links is nonconvex in the presence of
multiuser interference, the authors of \cite{Xiao04,Chiang07} considered
only the orthogonal wireless links which effectively reduced the problem to
convex one. In \cite{Neely05,Ribeiro10,Georgiadis06,Shroff06}, the interference
was considered in a fast fading environment but the proposed algorithms required
solving difficult subproblems. In \cite{Shao11}, the network was
approximated by a deterministic channel model \cite{Avestimehr11}
through which an approximate optimal solution was derived. A
similar joint optimization problem was also investigated in
\cite{Zheng12} for a wireless sensor network whereby a distributed algorithm
capable of converging to the stationary solution is
proposed. However, this approach is valid only for the setting with single
antenna nodes, and requires the utility function to be
strongly convex.

In this paper, we propose an efficient algorithm for joint backhaul
traffic engineering and physical layer interference management for
a large-scale SD-RAN. In particular, we leverage the Alternating
Direction Method of Multipliers (ADMM)
\cite{Bertsekas97,BoydADMMsurvey2011} and the WMMSE interference
management algorithm \cite{Razaviyayn13} to tackle the joint resource
allocation and traffic engineering problem. The resulting algorithm
is significantly more efficient than the subgradient-based methods \cite{Xiao04,Chiang07}.
The proposed algorithm has simple closed-form updates in each step and
is well suited for distributed and parallel implementation. Moreover, the proposed algorithm can
be extended to deal with per-flow quality of service constraint, or to networks with multi-antenna nodes.
Since not all the QoS requirements can be met simultaneously, techniques from sparse
optimization \cite{Yuan06,Candes08} are used to dynamically
select the subset of users being served. The
efficacy and the efficiency of the proposed algorithms are
demonstrated via extensive simulations.


{\it Notations}: We use ${\bf I}$ to denote the identity matrix, and
${\bf 0}$ to denote a zero vector or matrix. The superscripts `$T$',
`$H$' and `$*$', respectively, stand for the transpose, the
conjugate transpose and the complex conjugate. The indicator function for a
set $\mathcal{A}$ is denoted by $1_{\mathcal{A}}(x)$, that is,
$1_{\mathcal{A}}(x)=1$ if $x\in\mathcal{A}$, and
$1_{\mathcal{A}}(x)=0$ otherwise. The projection
function to the nonegative orthant is denoted by $(x)^{+}$, i.e., $(x)^{+}\triangleq
\max\{0,x\}$. Also, the notation $0\leq a\bot b\geq 0$ means that $a,b\geq 0$ and
$ab=0$. Some other notations are summarized in Table~\ref{tableSymbols}. \vspace*{-0.1cm}
\begin{table}
\caption{\small {A List of Notations} }\vspace*{-0.5cm}
\begin{center}
{\small
\begin{tabular}{|c |c | c| c| }
\hline
$\mathcal{V}$& The set of nodes in the network & $\mathcal{N}$& The set of routers\\
\hline
$\mathcal{B}$& The set of BSs  &$\mathcal{U}$& The set of mobile users\\
\hline
$\mathcal{L}$& The set of links  & $M$& Number of total commodities in the system\\
\hline
$\cL^{w}$& The set of wired links & $\cL^{wl}$& The set of wireless links\\
\hline
$C_{l}$& The capacity for a wired link $l\in \cL^{w}$& $K$& Number of tones on each wireless link \\
\hline
$r_m(l)$& Transmit rate for commodity $m$ on link $l$& $r_m$& Data rate for commodity $m$ \\
\hline
$D(m)$& The destination node for commodity $m$ & $S(m)$& The source node for commodity $m$\\
\hline
$p^k_{ds}$& The precoder from BS $s$ to user $d$ on tone $k$ & $I(l)$& The set of interferer to wireless link $l$\\
 \hline
\end{tabular} } \label{tableSymbols}
\end{center}
\vspace*{-0.2cm}
\end{table}

\vspace{-0.2cm}
\section{System Model and Problem Formulation}\label{Sec:Model}

Let $\mathcal{V}$ denote the set of nodes in a HetNet, comprised of
a set of network routers  $\mathcal{N}$, a set of BSs
$\mathcal{B}$, and a set of mobile users $\mathcal{U}$. Let $\mathcal{L}$
denote the set of
directed links that connect the nodes of $\mathcal{V}$.
In addition, we assume that there are $M$
source-destination pairs, denoted by $\{(S(m),D(m))\}_{m=1}^M$. For
each $m=1,...,M$, a data flow of rate $r(m)\ge0$ is to be sent from
the source node $S(m)$ to the destination node $D(m)$ over the
network.

The set of directed links $\cL$ consists of both wired and wireless
links. The wired links connect routers in
$\mathcal{N}$ and BSs in $\mathcal{B}$, and is denoted as
$\mathcal{L}^{w}\triangleq\{(s,d)\mid (s,d)\in\mathcal{L},~\forall\;
s,d\in\mathcal{N}\cup \mathcal{B}\}$. Here $(s,d)$ denotes the
directed link from node $s$ to node $d$. Assume each wired link
$l\in\mathcal{L}^{w}$ has a fixed capacity,
$C_{l}$. Then the total flow rate on link $l$ is constrained by
\vspace{-0.2cm}
\begin{align}\label{CapacityWired}
\sum_{m=1}^{M}r_{l}(m)\leq C_{l},~\forall\;l\in\mathcal{L}^{w},
\end{align}
where $r_{l}(m)\geq 0$ denotes the nonnegative flow rate on link $l$
for commodity $m$.

The wireless links provide single-hop connections between the
BSs to the mobile users. We assume that each BS divides the spectrum into
$K$ orthogonal frequency subchannels, and refer to these subchannels as {\it wireless links}.
Thus, the set of wireless links can
be represented as $$\mathcal{L}^{wl}\triangleq \{(s,d,k)\mid
(s,d,k)\in\mathcal{L},~\forall\; s\in\mathcal{B},~\forall\;
d\in\mathcal{U},~k=1\sim K\}$$ with $(s,d,k)$ being the wireless link
from node $s$ to node $d$ on subchannel $k$. For subchannel
$k$, BS $s\in\mathcal{B}$ applies a linear scalar precoder
$p_{ds}^{k}\in\Cplx$ to the transmitted complex unit-norm symbol of
mobile user $d\in\mathcal{U}$, so each mobile user can be served by
multiple BSs. Assuming that each mobile user treats the interference
from other BSs as noise, the total flow rate constraint on the wireless
link $l=(s,d,k)\in\mathcal{L}^{wl}$ is expressed as \vspace{-0.2cm}
\begin{align}\label{CapacityWireless}
\sum_{m=1}^{M}r_{l}(m)&\leq \bar r_{l}({\bf p})\triangleq
\log\left(1+\frac{|h_{ds}^{k}|^{2}|p_{ds}^{k}|^{2}}{\sum\limits_{(s',d',k)\in
I(l)\setminus\{l\}}|h_{ds'}^{k}|^{2}|p_{d's'}^{k}|^{2}+\sigma_{d}^{2}}\right),~\forall\;l\in\mathcal{L}^{wl},
\end{align}
where ${\bf p}\triangleq\{p_{ds}^{k}\mid \forall\;
(s,d,k)\in\mathcal{L}^{wl}\}$; $h_{ds}^{k}\in\Cplx$ is the
channel tap for the wireless link $l=(s,d,k)$; $\sigma_{d}^{2}$ is the
variance of AWGN noise at mobile user $d$;
$I(l)\subseteq\mathcal{L}^{wl}$ is the set of links interfering with link $l$:
\begin{align}
I(l)\triangleq\{(s',d',k)\in\mathcal{L}^{wl}\mid h^k_{ds'}\ne 0,
(s,d,k)=l\}.
\end{align}
Note that in this definition, link $l$ itself is included in $I(l)$, i.e., we have $l\in I(l)$.
Each BS $s\in\mathcal{B}$ has a total power budget $\bar
p_{s}\ge 0$, satisfying
\begin{align}\label{PowerConstraint}
\sum_{k=1}^{K}\sum_{d:(s,d,k)\in\mathcal{L}^{wl}}|p_{ds}^{k}|^{2}\leq
\bar p_{s},~\forall\;s\in\mathcal{B}.
\end{align}
Each node in the network should follow the flow conservation
constraint, i.e., the total incoming flow of node $v\in\mathcal{V}$
equals the total outgoing flow of that node,
\begin{align}\label{ConservationConst}
&\sum_{l\in{\rm In}(v)}r_{l}(m)+1_{\{S(m)\}}(v)r_{m}
=\sum_{l\in{\rm Out}(v)}r_{l}(m)+1_{\{D(m)\}}(v)r_{m}, \; m=1\sim
M,~\forall\; v\in\mathcal{V}
\end{align}
where ${\rm In}(v)$ and ${\rm Out}(v)$ denote the set of
links going into and coming out of a node $v$ respectively.

In this paper, we are interested in maximizing the minimum flow rate
of all commodities, while jointly performing the following tasks
{\it 1)}: route $M$ commodities from node $S(m)$ to node $D(m)$,
$m=1\sim M$; and {\it 2)} design the linear precoder at each BS.
This problem can be formulated as\vspace{-0.3cm}
\begin{align}\label{MainProb}
\max_{{\bf p},{\bf r}}&\quad r \\
{\rm s. t.}&\quad r\geq 0,~r_{m}\geq r,~r_{l}(m)\geq
0,~m=1\sim M,~\forall\;l\in\mathcal{L}\nonumber\\
&\quad\eqref{CapacityWired},~\eqref{CapacityWireless},~\eqref{PowerConstraint},~\mbox{and}~\eqref{ConservationConst}\nonumber,
\end{align}
where ${\bf r}\triangleq\{r,\;r_{l}(m),\;r_{m}\mid
\forall\;l\in\mathcal{L},~m=1\sim M\}$. Adopting the min-rate
utility results in a fair rate allocation, and such utility has been
adopted by many recent works in both the SDN and wireless
communities; see \cite{Razaviyayn13, Danna12} and the references
therein. At this point, it is important to note that by solving
problem \eqref{MainProb}, we automatically select a subset of BSs in
$\mathcal{B}$ to serve each user. That is, for a given commodity $m$
for user $d$, it is possible that there exist $r_{(s,d,k)}(m)>0$ and
$r_{(q, d, l)}(m)>0$ with $s\ne q$, and $(s,d,k), (q,d,l)\in
\mathcal{L}^{wl}$. Allowing cooperation among the BSs is in
agreement with the envisioned next generation cellular networks
\cite{HuaweiSDN}, which will rely heavily on  various  BS
cooperation schemes such as joint processing to improve the
transmission rate. Here, for simplicity, we don't take joint
processing between BSs into consideration.

Problem \eqref{MainProb} is difficult to solve because of the following reasons:
 \begin{itemize}
  \item [i)] It is a nonconvex problem where the nonconvexity comes from the rate constraints on the wireless links
  \item [ii)] The conventional approaches such as the bisection procedure for solving the max-min rate power allocation (beamformer) design \cite{Wiesel06} cannot be applied here, due to the existence of the conservation constraints and the presence of multiple frequency tones.
  \item [iii)] The size of the problem can be huge, as a result even if
we consider the simplest scenario in which there are no mobile users
(or equivalently the nonconvex wireless rate constraints are not
present), the resulting problem may still be difficult to solve in real time.
\end{itemize}
In the following, we propose an efficient distributed algorithm to compute a
stationary solution of the problem \eqref{MainProb}.

\vspace{-0.0cm}
\section{Joint Traffic Engineering and Interference Management}\label{Sec:Wireless}
\vspace{-0.0cm}

In this section, we propose a distributed algorithm that solves
problem \eqref{MainProb} to a stationary solution. We emphasize that
this problem is nonconvex due to the flow rate constraints on
wireless links, i.e., \eqref{CapacityWireless}.

\subsection{Algorithm Outline}

A special case of the considered problem model is known as $M$-pair
interference channel with the following settings i) the number of
BSs is the same as that of the users, i.e.,
$|\mathcal{B}|=|\mathcal{U}|=M$;  ii) there are only wireless links
for each wireless transmitter and user pair, i.e. $|\cN|=0$; iii)
each wireless transmitter and user pair serve, respectively, as the
source and the destination node of a commodity. For this special
case, it has been shown that the minimum rate maximization problem
is NP-hard when both transmitter and user are equipped with no less
than $2$ antennas \cite{Razaviyayn13}. However, when the wireless
transmitters are equipped with multiple antennas (resp. single
antenna) while mobile users are equipped with only one antenna
(resp. multiple antennas), the nonconvex minimum rate maximization
problem has been shown to be polynomial time solvable for the single
tone case of $K=1$ \cite{BK:Bengtsson01,Wiesel06, liu13spl}.
However, this is no longer true if there is more than one frequency
tone.

In the following, we will propose an efficient algorithm that can
solve problem \eqref{MainProb} to a stationary solution.
The proposed algorithm is a combination of two algorithms: 1) the max-min WMMSE
algorithm developed in \cite{Razaviyayn13} for minimum rate
maximization in $M$-pair interference channel; 2) the ADMM algorithm
that is used to distributively solve the multi-commodity routing
problem. Central to the proposed approach is the utilization of a
rate-MSE relationship, stated below  \cite{Razaviyayn13}.
\begin{lemma}\label{Rate-MSE}
For a given $l=(s,d,k)\in\mathcal{L}^{wl}$, $\bar r_{l}({\bf p})$
can be equivalently expressed as
\begin{align}\label{Rate-MSERelation}
\bar r_{l}({\bf p})=\max_{u_{l}, w_{l}} E_{l}(u_{l},w_{l},{\bf
p})\triangleq\max_{u_{l}, w_{l}}
c_{1,l}+c_{2,l}p_{ds}^{k}-\sum_{n=(s',d',k)\in
I(l)}c_{3,ln}|p_{d's'}^{k}|^{2}
\end{align}
where $(c_{1,l}, c_{2,l}, c_{3,ln})$ are given by $
c_{1,l}=1+\log(w_{l})-w_{l}(1+\sigma_{d}^2|u_{l}|^{2})$,
$c_{2,l}=2w_{l} {\rm Re}\{{u_{l}^{*}}h_{ds}^{k}\}$, and $
c_{3,ln}=w_{l}|u_{l}|^{2}|h_{ds'}^{k}|^{2}$.
\end{lemma}

Note that Lemma \ref{Rate-MSE} reformulates $\bar{r}_l(\mathbf{p})$
by introducing two extra sets of variables ${\bf
u}\triangleq\{u_{l}\mid l\in\mathcal{L}^{wl}\}$ and ${\bf
w}\triangleq\{w_{l}\mid l\in\mathcal{L}^{wl}\}$, with one pair of variables $\{u_l, w_l\}$ for each wireless link $l$. The term inside the
maximization operator is the MSE for estimating the message
transmitted on link $l$. Given Lemma \ref{Rate-MSE}, we reformulate
problem \eqref{MainProb} by replacing $\bar r_{l}({\bf p})$ in
\eqref{MainProb} with its MSE. We call such new constraint a {\it
rate-MSE constraint}. Then, we consider the following problem with two extra
optimization variable sets ${\bf u}$ and ${\bf w}$  instead: \vspace{-0.0cm}
\begin{align}\label{WMMSEQ}
\max~&r\\
{\rm s. t.}~&r\geq 0,~r_{m}\geq r,~r_{l}(m)\geq
0,~m=1\sim M,~\forall\;l\in\mathcal{L},\nonumber\\
&\eqref{CapacityWired},~\eqref{PowerConstraint}, \mbox{~and~}
\eqref{ConservationConst},\nonumber\\
&\sum_{m=1}^{M}r_{l}(m)\leq
c_{1,l}+c_{2,l}p_{ds}^{k}-\sum_{n=(s',d',k)\in
I(l)}c_{3,ln}|p_{d's'}^{k}|^{2},\;\forall\;l\in\mathcal{L}^{wl}.\label{RateMSEConstraint}
\end{align}
Why do we include these extra optimization variables ${\bf u}$ and ${\bf w}$? First we observe that for any given $\{{\bf r},{\bf p}\}$, the optimal
${\bf u}$ (resp. ${\bf w}$) for \eqref{Rate-MSERelation} can be
obtained while ${\bf w}$ (resp. ${\bf u})$ is held fixed. Moreover,
these optimal solutions can be expressed in closed form for any
$l\in\mathcal{L}^{wl}$:
\begin{align}
u_{l}&=\bigg(\sum_{(s',d',k)\in I(s,d,k)}\mid h_{ds'}^{k}|^{2}|p_{d's'}^{k}|^{2}+\sigma_{d}^{2}\bigg)^{-1}h_{ds}^{k}p_{ds}^{k},\label{UpdateU}\\
w_{l}&=\bigg(1-(h_{ds}^{k}p_{ds}^{k})^{*}u_{l}\bigg)^{-1}.\label{UpdateW}
\end{align}
These expressions suggest that the set of variables ${\bf u}$ and
${\bf w}$ can be updated independently and locally at each mobile
user if the interference plus noise and local channel state
information are locally known to the users. Moreover, when $\bf
u$ and $\bf w$ are fixed, the problem for updating $\{{\bf
r},{\bf p}\}$ is convex (note that \eqref{Rate-MSERelation} is a
convex quadratic problem on the precoders $\mathbf{p}$) and
can be solved in polynomial time. Hence, we propose to apply the alternating
optimization technique to solve problem \eqref{WMMSEQ}; see the
N-MaxMin Algorithm in Table~\ref{NMaxMinAlgorithm} for a detailed
description.

The following result states that the iterates
$\{{\bf r}^{(t)},{\bf p}^{(t)}\}$ generated by the above algorithm
converge to a stationary solution of the original problem
\eqref{MainProb}. The proof of this result is relegated to
Appendix \ref{Appendix:WMMSEProof}.

\begin{Theorem}\label{BCDConverge}
The sequence $\{{\bf r}^{(t)},{\bf p}^{(t)}\}$ generated by the
N-MaxMin Algorithm converges to a stationary solution of problem
\eqref{MainProb}. Moreover, every global optimal solution of problem
\eqref{MainProb} corresponds to a global optimal solution of the
reformulated problem \eqref{WMMSEQ}, and they achieve the same
objective value.
\end{Theorem}\vspace{-0.1cm}

\begin{table}[t]
\begin{center}
\fbox{\parbox[]{.95\linewidth}{ \vspace{0cm} \noindent {\bf Network
Max-Min WMMSE (N-MaxMin) Algorithm}:
\begin{algorithmic}[1]
\State {\bf Initialization} Generate a feasible set of variables
$\{{\bf r},{\bf p}\}$, and let $t=1$.

\State {\bf Repeat}

\State ~~~${\bf u}^{(t)}$ is updated by \eqref{UpdateU}

\State ~~~${\bf w}^{(t)}$ is updated by \eqref{UpdateW}

\State ~~~$\{{\bf r}^{(t)},{\bf p}^{(t)}\}$ is updated by solving
the problem \eqref{WMMSEQ} via Algorithm 1 in Table
\ref{Algorithm1}

\State ~~~$t=t+1$

\State{\bf Until} Desired stopping criteria is met

\end{algorithmic}
}}
\end{center}\vspace{-0.1cm}
\caption{Network
Max-Min WMMSE (N-MaxMin) Algorithm}\label{NMaxMinAlgorithm}\vspace{-1cm}
\end{table}

\begin{Rmk} The N-MaxMin Algorithm (Table~\ref{NMaxMinAlgorithm}) and its
convergence analysis (Theorem \ref{BCDConverge}) extend easily
to the multi-antenna case. The key is to use the matrix version of Lemma \ref{Rate-MSE} in \cite{Razaviyayn13}.
\end{Rmk}

\subsection{A Brief Review of ADMM Algorithm}

The second ingredient for the proposed approach is to use the ADMM
algorithm to update $\{{\bf r},{\bf p}\}$ in the N-MaxMin Algorithm.
Unlike the computation of ${\bf u}$ and ${\bf w}$, the updates for
$\{{\bf r},{\bf p}\}$ do not have closed forms. We can use off-the-shelve toolboxes, but this is not
very efficient. In the sequel, we first use variable splitting
to decompose the problem and then solve it using ADMM. The resulting algorithm
has closed form updates in each step and
is well suited for parallel and distributed implementation.

We now briefly review the ADMM algorithm. Consider the following structured
convex problem \cite{BoydADMMsurvey2011},\vspace{-0.3cm}
\begin{align}\label{ADMM}
\min_{{\bf x}\in\Cplx^{n},{\bf z}\in\Cplx^{m}}~& f({\bf x})+g({\bf z})\notag\\
{\rm s.t.}~& {\bf A}{\bf x}+{\bf B}{\bf z}={\bf c}\\
& {\bf x}\in \mathcal{C}_{1},~{\bf z}\in\mathcal{C}_{2}\notag
\end{align}
where ${\bf A}\in\Cplx^{k\times n}$, ${\bf B}\in\Cplx^{k\times m}$,
${\bf c}\in\Cplx^{k}$; $f$ and $g$ are convex functions;
$\mathcal{C}_{1}$ and $\mathcal{C}_{2}$ are non-empty convex
sets. The partial augmented Lagrangian function for problem \eqref{ADMM}
can be expressed as
\begin{align}\label{ADMMStructure}
L_{\rho}({\bf x},{\bf z}, {\bf y})=f({\bf x})+g({\bf z})+{\rm
Re}\left({\bf y}^{H}({\bf A}{\bf x}+{\bf B}{\bf z}-{\bf
c})\right)+(\rho/2)\|{\bf A}{\bf x}+{\bf B}{\bf z}-{\bf c}\|_2^2
\end{align}
where ${\bf y}\in\Cplx^{k}$ is the Lagrangian dual variable
associated with the linear equality constraint, and $\rho>0$ is some
constant. The ADMM algorithm solves problem \eqref{ADMM} by
iteratively performing three steps in each iteration $t$:
\begin{subequations}
\begin{align}
{\bf x}^{(t)}&=\arg\min_{{\bf x}\in\cC1}L_{\rho}({\bf x},{\bf
z}^{(t-1)},{\bf
y}^{(t-1)}) \quad \mbox{(primal update for the first block variable)}\label{ADMMUpdate1}\\
{\bf z}^{(t)}&=\arg\min_{{\bf z}\in\cC2}L_{\rho}({\bf x }^{(t)},{\bf z},{\bf
y}^{(t-1)})\quad ~~~\mbox{(primal update for the second block variable)}\label{ADMMUpdate2}
\\
{\bf y}^{(t)}&={\bf y}^{(t-1)}+\rho({\bf A}{\bf x}^{(t)}+{\bf B}{\bf
z}^{(t)}-{\bf c})\quad ~\mbox{(dual variable update)}\label{ADMMUpdate3}.
\end{align}
\end{subequations}
The practical efficiency of ADMM can be attributed to the fact that in many
applications, the subproblems \eqref{ADMMUpdate1} and \eqref{ADMMUpdate2} are solvable in closed-form.
The convergence and the optimality of the algorithm is summarized in the following lemma \cite{Bertsekas97}.
\begin{lemma}\label{ADMMConverge} Assume that the optimal solution set of problem \eqref{ADMM} is non-empty, and ${\bf A}^{T}{\bf A}$ and ${\bf B}^{T}{\bf B}$ are invertible. Then the sequence of $\{{\bf x}^{(t)}, {\bf z}^{(t)},{\bf y}^{(t)}\}$ generated by \eqref{ADMMUpdate1}, \eqref{ADMMUpdate2}, and \eqref{ADMMUpdate3} is bounded and every limit point of $\{{\bf x}^{(t)},{\bf z}^{(t)}\}$ is an
optimal solution of problem \eqref{ADMM}.
\end{lemma}

\subsection{An ADMM Approach for Updating $\{{\bf r},{\bf p}\}$}\label{Sec:WirelessUpdate}

In the following, we will first reformulate the subproblem for $\{{\bf r},{\bf p}\}$ into the form of \eqref{ADMM}, so that the ADMM can be applied. Then we will show that each step of the resulting algorithm is easily computable and amenable for distributed implementation. To this end, we will appropriately split the variables in the coupling constraints \eqref{ConservationConst} and \eqref{RateMSEConstraint}.

We first observe that each flow rate $r_{l}(m)$ on link $l=(s,d)\in\mathcal{L}^{w}$ (or $l=(s,d,k)\in \mathcal{L}^{wl}$) for commodity $m$ is shared among {\it two} flow conservation constraints, one for node $s$ and the other for node $d$.  To induce separable subproblems and enable distributed computation,
we introduce two local auxiliary copies of $r_l(m)$, namely $\hat
r_{l}^{s}(m)$ and $\hat r_{l}^{d}(m)$, and store one at node $s$ and the other at node $d$. Similarly, we introduce two local auxiliary copies for each commodity rate, denoted as $\hat r_{m}^{S(m)}$, $\hat r_{m}^{D(m)}$, $m=1\sim M$, and store them at the source and the destination node of each commodity, respectively.
That is, we have introduced the following auxiliary variables:
\begin{subequations}\label{SlackLinks}
\begin{align}
&\hat r_{m}^{S(m)}=r_{m}, \; \hat{r}_{m}^{D(m)}=r_{m}, \; m=1\sim M; \label{SlackRate}\\
&\hat r_{l}^{s}(m)=r_{l}(m), \; \hat r_{l}^{d}(m)=r_{l}(m),\;
\forall\;l=(s,d)\in\mathcal{L}^{w}; \label{SlackWired}\\
&\hat r_{l}^{s}(m)=r_{l}(m), \; \hat r_{l}^{d}(m)=r_{l}(m),\;
\forall\;l=(s,d,k)\in\mathcal{L}^{wl}.\label{SlackWireless}
\end{align}
\end{subequations}

The flow rate
conservation constraints on node $v\in\mathcal{V}$ can then be
rewritten as
\begin{align}
&\sum_{l\in{\rm In}(v)}\hat r_{l}^{v}(m)+1_{\{S(m)\}}(v)\hat
r_{m}^{v}=\sum_{l\in{\rm Out}(v)}\hat
r_{l}^{v}(m)+1_{\{D(m)\}}(v)\hat r_{m}^{v},~m=1\sim
M.\label{ConservationConstNew}
\end{align}\vspace{-0.2cm}

In addition, for the rate-MSE constraint, we introduce several
copies of the transmit precoder on a given wireless link
$l=(s,d,k)\in\mathcal{L}^{wl}$, i.e.
\begin{align}
p_{d's',ds}^{k}=p_{ds}^{k},\; \forall\;l\in
I(s',d',k)\subset\mathcal{L}^{wl}. \label{SlackPrecoder}
\end{align}

Intuitively, by doing such variable splitting,
each variable $p_{d's',ds}^{k}$ will only appear in {\it a single}
rate-MSE constraint. For a given link
$l=(s,d,k)\in\mathcal{L}^{wl}$, its rate-MSE constraint only depends
on the set of precoders $\{p_{ds,d's'}^{k}\mid \forall\;(s',d',k)\in
I(l)\}$, as can be seen below\vspace{-0.2cm}
\begin{align}
\sum_{m=1}^{M}r_{l}(m)\leq
c_{1,l}+c_{2,l}p_{ds,ds}^{k}-\hspace{-0.6cm}\sum_{n=(s',d',k)\in
I(l)}c_{3,ln}|p_{ds,d's'}^{k}|^{2},~\forall\;l\in\mathcal{L}^{wl}.\label{Rate-MSENew}\vspace{-0.3cm}
\end{align}
Moreover, for the analysis of the convergence result, another auxiliary
variable $\hat r$ is introduced such that $r=\hat r$.

Using these new variables, the updating step for $\{{\bf r},{\bf p}\}$ is equivalently
expressed as\vspace{-0.2cm}
\begin{align}\label{WMMSEQEquiv}
\max ~&(r+\hat r)/2\nonumber\\
{\rm s. t.}~&r=\hat r,~r\geq 0,~r_{m}\geq r,~r_{l}(m)\geq 0, ~m=1\sim M,~l\in\mathcal{L}\nonumber\\
&\eqref{CapacityWired},~\eqref{PowerConstraint},~\eqref{SlackLinks},~\eqref{ConservationConstNew},~\eqref{SlackPrecoder}~\mbox{and}~\eqref{Rate-MSENew}.
\end{align}
It is important to note that the constraints of problem
\eqref{WMMSEQEquiv} (except the linear equality constraints $r=\hat
r$, \eqref{SlackLinks} and \eqref{SlackPrecoder}) are now separable
between two optimization variable sets {\it i)} the tuple $\{{\bf
r},\hat{\bf p}\}$ where $\hat{\bf p}\triangleq\{p_{d's',ds}^{k}\mid
\forall\;l=(s,d,k),\;l'=(s',d',k)\in\mathcal{L}^{wl},~l\in I(l')\}$,
and {\it ii)} the tuple $\{\hat{\bf r},{\bf p}\}$ where $\hat{\bf
r}\triangleq\left\{\hat r,\hat r_{m}^{S(m)},\hat r_{m}^{D(m)},\hat
r_{l}^{s}(m),\hat r_{l}^{d}(m)\mid m=1\sim M,~\forall
\;l=(s,d)~\mbox{or}~(s,d,k)\in\mathcal{L}\right\}$. Additionally,
the objective function is linear and separable over $\rb$ and
$\hat\rb$. Therefore the ADMM algorithm can be used to solve problem
\eqref{WMMSEQEquiv}. The resulting algorithm, described in
Table~\ref{Algorithm1}, is referred to as Algorithm 1. Note that the
partial augmented Lagrange function for problem \eqref{WMMSEQEquiv}
is given by {\small
\begin{align*}
&L_{\rho_{1},\rho_{2}}({\bf r},\hat{\bf p},\hat{\bf r},{\bf p};\bfdelta,\bftheta)=(r+\hat r)/2+\delta(\hat r-r)-\frac{\rho_{1}}{2}(\hat r-r)^{2}\nonumber\\
&+\underbrace{\sum_{m=1}^{M}\left[\delta_{m}^{S(m)}(\hat
r_{m}^{S(m)}-r_{m})+\delta_{m}^{D(m)}(\hat
r_{m}^{D(m)}-r_{m})-\frac{\rho_{1}}{2}(\hat
r_{m}^{S(m)}-r_{m})^{2}-\frac{\rho_{1}}{2}(\hat
r_{m}^{D(m)}-r_{m})^{2}\right]}_{\mbox{enforcing linear constraints
\eqref{SlackRate}}}+\sum_{m=1}^{M}\\
&\sum_{l=(s,d)\in\mathcal{L}\atop{l=(s,d,k)\in
\mathcal{L}^{wl}}}\underbrace{\left[\delta_{l}^{s}(m)(\hat
r_{l}^{s}(m)-r_{l}(m))+\delta_{l}^{d}(m)(\hat
r_{l}^{d}(m)-r_{l}(m))-\frac{\rho_{1}}{2}(\hat
r_{l}^{s}(m)-r_{l}(m))^{2}
-\frac{\rho_{1}}{2}(\hat r_{l}^{d}(m)-r_{l}(m))^{2}\right]}_{\mbox{enforcing linear constraints \eqref{SlackWired} and \eqref{SlackWireless}}}\notag\\
&+\underbrace{\sum_{l=(s,d,k)\in\mathcal{L}^{wl}\atop{n=(s',d',k)\in
I(s,d,k)}}\left[\theta_{ln}(p_{d's'}^{k}-p_{ds,d's'}^{k})-\frac{\rho_{2}}{2}(p_{d's'}^{k}-p_{ds,d's'}^{k})^{2}\right]}_{\mbox{enforcing
linear constraints \eqref{SlackPrecoder}}},
\end{align*}}
where we have used $\delta$, $\{\delta_{m}^{S(m)}\}$,
$\{\delta_{m}^{D(m)}\}$, $\{\delta_{l}^{s}(m)\}$,
$\{\delta_{l}^{d}(m)\}$ and $\{\theta_{ln}\}$ to denote the
Lagrangian multipliers for various equality constraints, and have
collected these multipliers to the vectors $\bfdelta$ and
$\bftheta$; $\rho_{1}>0$ and $\rho_{2}>0$ are some constant
coefficients for, respectively, the linear equality constraints
\eqref{SlackLinks} and \eqref{SlackPrecoder}. For notational
simplicity, let us stack all the elements of $\mathbf{r}$ and
$\hat{\mathbf{r}}$ to the following vectors
\begin{align*}
&\rb_{\rm stack}\triangleq[r,\{r_{m}\}_{m=1\sim M},\{r_{l}(m)\}_{m=1\sim M,l\in\mathcal{L}}]^{T}\\
&\hat\rb_{\rm stack}\triangleq[\hat r,\{r_{m}^{S(m)}\}_{m=1\sim M},\{r_{m}^{D(m)}\}_{m=1\sim M},\{r_{l}^{s}(m)\}_{m=1\sim M,l=(s,d)\in\mathcal{L}},\{r_{l}^{d}(m)\}_{m=1\sim M,l=(s,d)\in\mathcal{L}}]^{T}.
\end{align*}
Similarly, stack all the elements of $\mathbf{p}$ and
$\hat{\mathbf{p}}$ by
\begin{align*}
&\mathbf{p}_{\rm stack}\triangleq\{p^{k}_{ds}, \; \forall~(s,d,k)\in \mathcal{L}^{wl}\}\nonumber\\
&\hat{\mathbf{p}}_{\rm stack}\triangleq\left\{\{p^{k}_{d's', ds}, \forall\; (s,d,k)\in I(s',d',k')\}, \forall\; (s,d,k)\in \mathcal{L}^{wl}\right\}.
\end{align*}

Then the equality relationships \eqref{SlackRate}--\eqref{SlackWireless} and \eqref{SlackPrecoder} can be compactly expressed
as\vspace{-0.0cm}
\begin{align}\label{Dummy}
\mathbf{C}{\bf r}_{\rm stack}=\hat{\bf r}_{\rm stack},\; \mathbf{D}{\bf p}_{\rm stack}=\hat{\bf p}_{\rm stack},
\end{align}
where
\begin{align}
\mathbf{C}=\left[\begin{array}{ccccc}
1&0&0&0&0\\
0&{\bf I}&{\bf I}&{\bf 0}&{\bf 0}\\
0&{\bf 0}&{\bf 0}&{\bf I}&{\bf I}
\end{array}\right]^{T};\quad \mathbf{D}={\rm blkdg}[\{\mathbf{1}^k_{ds}\}_{(s,d,k)\in \mathcal{L}^{wl}}]
\end{align}
where $\rm blkdg\{\cdot\}$ is the block diagonalization operator;
$\mathbf{1}^k_{ds}$ is an all one column  vector of size equal to
the total number of links with which $l=(s,d,k)$ interferes, given by $$|\bar
I(l)|\triangleq|\{(s',d',k)\mid (d,s,k)\in I(s',d',k)\}|.$$ Using the
notation in \eqref{Dummy}, we can simplify the above expression to
\begin{align*}
L_{\rho_{1},\rho_{2}}({\bf r},\hat{\bf p},\hat{\bf r},{\bf p};\bfdelta,\bftheta)&=r+\left[\bfdelta^{T}(\hat{\bf r}_{\rm stack}-\mathbf{C}{\bf r}_{\rm stack})-\frac{\rho_{1}}{2}\|\hat{\bf r}_{\rm stack}-\mathbf{C}{\bf r}_{\rm stack}\|^{2}\right]\nonumber\\
&\quad +\left[\bftheta^{H}(\mathbf{D}{\bf p}_{\rm stack}-\hat{\bf
p}_{\rm stack})-\frac{\rho_{2}}{2}\|\mathbf{D}{\bf p}_{\rm stack}-\hat{\bf p}_{stack}\|^{2}\right].
\end{align*}
Moreover, by appealing to the standard analysis for ADMM algorithm (Lemma \ref{ADMMConverge}), and using the fact that $\mathbf{C}^T\mathbf{C}$ and $\mathbf{D}^T\mathbf{D}$ are both full
rank matrices, we easily see that Algorithm 1 converges to the optimal solutions of problem \eqref{WMMSEQEquiv}.

\begin{figure*}[ht]
    \centering
     {\includegraphics[width=
0.7\linewidth]{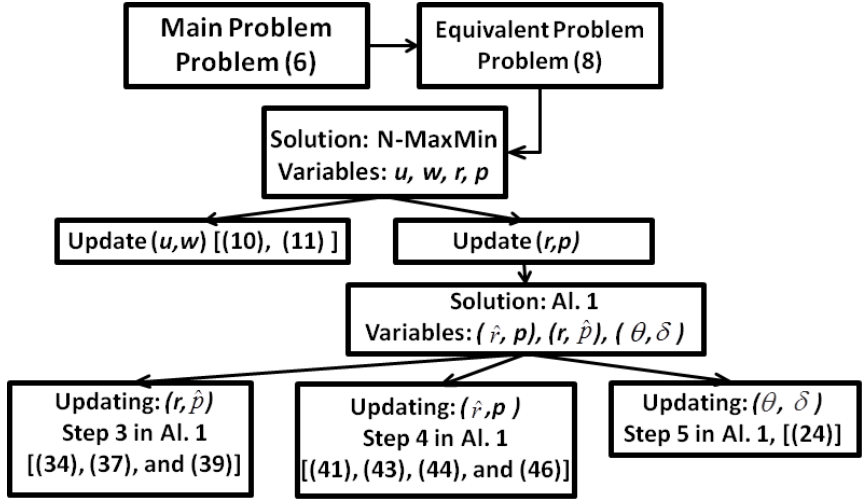}}\caption{Flow chart of the
proposed solution approach \eqref{MainProb}.
}\label{FigFlowChartProblem}
    \end{figure*}

For the detailed step-by-step specification of Algorithm 1, we refer the readers to Appendix \ref{Sec:StepByStep}.
The main message from the derivation therein is that each step in Algorithm 1 can be computed distributedly in closed-form. More specifically, Step 3 of the algorithm is decomposable {\it among all links} in the system (wireless and wired), while Step 4 of the algorithm is decomposable {\it among all
the nodes} in the system (also see Section \ref{Sec:Distributed} for elaboration).  These properties allow the entire algorithm to be easily implemented in a parallel fashion.  Fig.~\ref{FigFlowChartProblem} provides a flow chart showing the relationship among different algorithms

\begin{table}[t]
\begin{center}
\fbox{\parbox[]{.95\linewidth}{ \noindent {\bf Algorithm 1: ADMM for
\eqref{WMMSEQEquiv}}:
\begin{algorithmic}[1]
\State {\bf Initialize} all primal variables ${\bf r}^{(0)},\hat{\bf
r}^{(0)},{\bf p}^{(0)},\hat{\bf p}^{(0)}$ (not necessarily a
feasible solution for problem \eqref{WMMSEQEquiv}); Initialize all
dual variables $\bfdelta^{(0)},\bftheta^{(0)}$; set $t=0$

\State {\bf Repeat}

\State ~~~Solve the following problem and obtain ${\bf
r}^{(t+1)},\hat{\bf p}^{(t+1)}$:
\begin{align}\label{WMMSEQFirst}
\max_{{\bf r},\hat{\bf p}}~&
L_{\rho_{1},\rho_{2}}({\bf r},\hat{\bf p},\hat{\bf r}^{(t)},{\bf p}^{(t)};\bfdelta^{(t)},\bftheta^{(t)})\notag\\
{\rm s. t.}~&r\geq 0, r_{m}\geq r,~r_{l}(m)\geq 0, ~m=1\sim
M,~l\in\mathcal{L},\nonumber\\
&\eqref{CapacityWired}~\mbox{and}~\eqref{Rate-MSENew}
\end{align}
This step can be {\it solved in parallel across all links}, cf.
\eqref{UpdateFirstSub1Sol}, \eqref{UpdateFirstSub2Sol}, and
\eqref{WMMSEQFirstSub2Sol}.

\State ~~~Solve the following problem and obtain $\hat {\bf
r}^{(t+1)},{\bf p}^{(t+1)}$:
\begin{align}\label{WMMSEQSecond}
\max_{\hat{\bf r},{\bf p}}~&
L_{\rho_{1},\rho_{2}}({\bf r}^{(t+1)},\hat{\bf p}^{(t+1)},\hat{\bf r},{\bf p};\bfdelta^{(t)},\bftheta^{(t)})\notag\\
{\rm s.t.}~&\eqref{PowerConstraint}~\mbox{and}~\eqref{ConservationConstNew}
\end{align}
This problem can be {\it solved in parallel across all nodes}, cf.
\eqref{UpdateSecondSub1Sol}, \eqref{UpdateSecondIndividualSol},
\eqref{UpdateSecondIndividualSol2}, and \eqref{WMMSEQSecondSubSol}.

\State ~~~Update the Lagrange dual multipliers $\bfdelta^{(t+1)}$
and $\bftheta^{(t+1)}$ by
\begin{align}\label{UpdateDual}
\bfdelta^{(t+1)}&=\bfdelta^{(t)}-\rho_{1}(\hat{\bf  r}_{\rm stack}^{(t+1)}-\mathbf{C}{\bf r}_{\rm stack}^{(t+1)}),\notag\\
\bftheta^{(t+1)}&=\bftheta^{(t)}-\rho_{2}(\mathbf{D}{\bf
p}_{\rm stack}^{(t+1)}-\hat{\bf p}_{\rm stack}^{(t+1)}).
\end{align}

\State ~~~$t=t+1$

\State{\bf Until} Desired stopping criterion is met

\end{algorithmic}
}}
\end{center}\vspace{-0.1cm}
\caption{Summary of the proposed Algorithm 1}\label{Algorithm1}\vspace{-0.3cm}
\end{table}

\section{Distributed Implementation and Extensions}\label{Sec:WirelessQoS}

\subsection{Distributed Implementation and Information Exchange}\label{Sec:Distributed}
In this section, we briefly elaborate how the N-MaxMin algorithm can
be implemented in a distributed manner. Let us first look at the
implementation for the backhaul network (i.e., the update for
$\mathbf{r}$ and $\hat{\mathbf{r}}$ when ignoring the wireless
links). Suppose there is a master node in the system. Consider the
update of the optimization variable ${\bf r}$ in Step 3 of Algorithm
1 (cf.\ Step (i) in Appendix \ref{Sec:StepByStep}-1).  In this step, to
update $\{r,r_{m}\mid m=1\sim M\}$, the source node and destination
node of each commodity $m$, $m=1\sim M$, should respectively send
$\left(\hat r_{m}^{S(m)}-\frac{\delta_{m}^{S(m)}}{\rho_{1}}\right)$
and $\left(\hat
r_{m}^{D(m)}-\frac{\delta_{m}^{D(m)}}{\rho_{1}}\right)$ to the
assumed master node. After the master node applies
\eqref{UpdateFirstSub1Sol} to update $\{r,r_{m}\mid m=1\sim M\}$, it
would transmit $r_{m}$ back to node $S(m)$ and $D(m)$. To update
$r_{l}(m),~m=1\sim M,~\forall~l\in\mathcal{L}$, the procedure is
decoupled across {\it each link} (cf.\ step (ii) in Appendix
\ref{Sec:StepByStep}-1). Therefore without loss of
generality, we can let the destination node of each link
$l=(s,d)\in\mathcal{L}$ perform the bisection updating step
\eqref{UpdateFirstSub2Sol}. Thus, the source node of link $l$ should
transmit $M$ real values, $(\hat
r_{l}^{s}(m)-\frac{\delta_{l}^{s}(m)}{\rho_{1}})$, $\forall \;
m=1\sim M$, to the destination node of that link. After updating
$r_{l}(m),~m=1\sim M$, the destination node of the link would
transmit them back to the source node. After $\bf r$ is computed,
the second block variables $\hat{\bf r}$ and the Lagrange dual
variable $\bfdelta$ can be updated in each node, see
\eqref{UpdateSecondSub1Sol}, \eqref{UpdateSecondIndividualSol},
\eqref{UpdateSecondIndividualSol2}, and \eqref{UpdateDual}.

Next we discuss the implementation for the wireless part, i.e., the update for $\mathbf{p}$ and $\hat{\mathbf{p}}$. We assume that {\it i)} each mobile user has local channel state information from all interfering BSs;
and {\it ii)} $u_{l}$ and $w_{l}$ are updated according to \eqref{UpdateU} and
\eqref{UpdateW} respectively at the receiver side of link
$l\in\mathcal{L}^{wl}$. 
Let us first look at the update for $\hat{\bf p}\cup \{r_{l}(m)\mid
m=1\sim M,\;\forall\; l\in\mathcal{L}^{wl}\}$ (cf.\
\eqref{WMMSEQFirstSub2}). Recall that this step is decoupled over
each wireless link, and all necessary information needed for the
computation (such as ${\bf u}$, ${\bf w}$, ${\bf p}$ and the channel
state information) is available at each user except
$(r_{l}^{s}(m)-\frac{\delta_{l}^{s}(m)}{\rho_{1}})$, $m=1\sim M$. It
follows that this update can be processed at the mobile users $d$,
provided that for wireless link $l=(s,d,k)\in\mathcal{L}^{wl}$, the
BS $s$ sends $(r_{l}^{s}(m)-\frac{\delta_{l}^{s}(m)}{\rho_{1}})$,
$m=1\sim M$ to mobile user $d$. After mobile user $d$ updates
$r_{l}(m),~m=1\sim M$, it sends them back to BS $s$. Next we analyze
the step that update $\mathbf{p}$ (cf. \eqref{WMMSEQSecondSub}). In
order to solve this problem locally at each BS $s\in\mathcal{B}$,
the mobile users whose transmissions interfere with the users
associated with BS $s$, i.e.,
\begin{align}
d'\in\left\{d'\mid (s',d',k)\in \bar
I(s,d,k), \forall \ d,~k=1\sim K, \mbox{ s.t.
}(s,d,k)\in\mathcal{L}^{wl}\right\}
\end{align}
should send
$(p_{d's',ds}^{k}+\frac{\theta_{(s',d'k),(s,d,k)}}{\rho_{2}})$, $\
\forall\  (s',d',k)\in\mathcal{L}^{wl}$ with BS $s$. After BS $s$
obtains the updated $p_{ds}^{k}$ by \eqref{WMMSEQSecondSubSol}, it
can broadcast these quantities back to those mobile users.

Given the information exchanges described above,
Algorithm 1 (and therefore, the N-MaxMin Algorithm) can be
implemented in a distributed and parallel manner. 

\subsection{Extension with Per-user QoS Requirements}
For a subset $\cal{Q}\subseteq$$\{ 1,\ldots,M\}$ of the end-to-end commodity
pairs, we may require the flow rates to be no less than
$\underline{r}_{q}$. For the rest of the commodities
$\mathcal{Q}^{c}\triangleq\{1,\ldots,M\}\setminus\mathcal{Q}$, we can maximize their
minimum achievable rate. This gives rise to the following formulation:
\begin{align}\label{MainProbQoS}
\max~&r\notag\\
{\rm s.\ t.}~&r\geq0,~r_{l}(m)\geq
0,~m=1\sim M,~\forall\;l\in\mathcal{L},\nonumber\\
&r_{q}\geq \underline{r}_{q},~\forall\; q\in\mathcal{Q},~r_{m}\geq r,~\forall\; m\in \mathcal{Q}^{c},\\
&\eqref{CapacityWired},~\eqref{CapacityWireless},~\eqref{PowerConstraint}, \mbox{~and~} \eqref{ConservationConst}.\notag
\end{align}

Different from problem \eqref{MainProb}, this QoS constrained formulation is not always feasible for any given  tuple of QoS constraints $\{\bar{r}_q\}_{q\in\cQ}$. Therefore, the N-MaxMin algorithm proposed in Table \ref{NMaxMinAlgorithm} cannot be directly applied.
To circumvent this difficulty, we introduce an extra optimization
variable set $$\bfalpha\triangleq\{\alpha_{q}\geq0\mid
q\in\mathcal{Q}\}.$$ The variable $\alpha_{q}$ can be interpreted as
the QoS violation for the $q$th QoS constraint. Using this set of
new variables, we replace the ``hard" QoS constraint
$r_{q}\geq\underline{r}_{q},\ \forall\ q\in\mathcal{Q}$ with the
following set of ``soft" constraints
$$ r_{q}\geq\underline{r}_{q}-\alpha_{q},~\forall
q\in\mathcal{Q}.$$
In this way problem \eqref{MainProbQoS} is always feasible. Hence, our goal becomes one that
selects the {\it maximum number} of commodities in the set $\cQ$ to satisfy the QoS requirements, in addition to
the joint optimization for power allocation and routing. In another word, besides optimizing $\mathbf{p}$ and $\mathbf{r}$, we would like to find a vector $\bfalpha$ that has the maximum number of zeros.

Mathematically, to induce zeros in $\bfalpha$, an extra regularization term that penalizes the nonzero terms in $\bfalpha$ should be added to the objective function of problem \eqref{MainProbQoS}: $
\max~r-\|{\bfalpha}\|_{0}$. Here the $\ell_0$ norm measures the number of nonzero elements
within a vector.  Follow the conventional sparse
optimization strategy \cite{Yuan06,Candes08}, we then relax the difficult $\ell_0$ norm to the convex $\ell_1$ norm, and consider the following problem instead
\begin{align}\label{MainProbQoSSparse}
\max~&r-\sum_{q\in\mathcal{Q}}\alpha_{q}\notag\\
{\rm s. t.}~&r\geq0,~r_{l}(m)\geq
0,~m=1\sim M,~\forall\; l\in\mathcal{L},\nonumber\\
&\bfalpha\geq 0, \ r_{q}+\alpha_{q}\geq \underline{r}_{q},\forall\; q\in\mathcal{Q},~r_{m}\geq r,~\forall\; m\in \mathcal{Q}^{c}\\
&\eqref{CapacityWired},~\eqref{CapacityWireless},~\eqref{PowerConstraint}, \mbox{~and~} \eqref{ConservationConst}.\notag
\end{align}
This problem can be solved to a stationary solution by applying a modified N-MaxMin algorithm. In particular,
the block variables are ${\bf u}$, ${\bf w}$, and $\{{\bf r},{\bf p},\bfalpha\}$. We observe that the updating procedures
for ${\bf u}$ and ${\bf w}$ are the same as in \eqref{UpdateU} and
\eqref{UpdateW}. To update $\{{\bf r},{\bf
p},\bfalpha\}$, we can apply the ADMM algorithm developed in Sec.\
\ref{Sec:WirelessUpdate} for problem \eqref{WMMSEQEquiv} with a few
minor modifications (omitted here due to space limitations).

\vspace{-0.0cm}
\section{Simulation Results}\label{Sec:Simulation}
\vspace{-0.1cm}

\begin{figure}[t]
\begin{center}
{ \subfigure[] []{\resizebox{.45\textwidth}{!}{\includegraphics{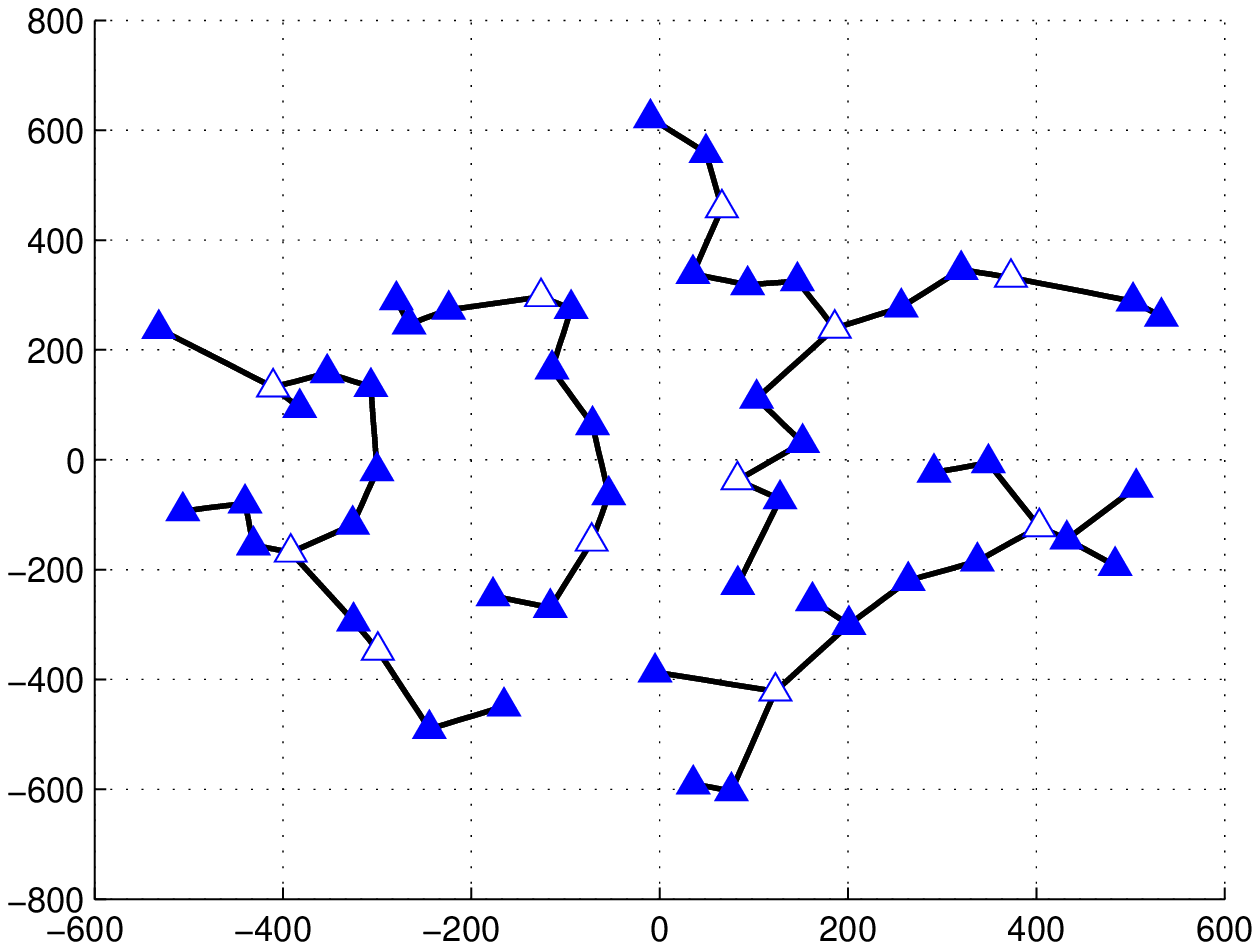}}}}
\hspace{1pc} { \subfigure[][ ]{\resizebox{.45\textwidth}{!}{\includegraphics{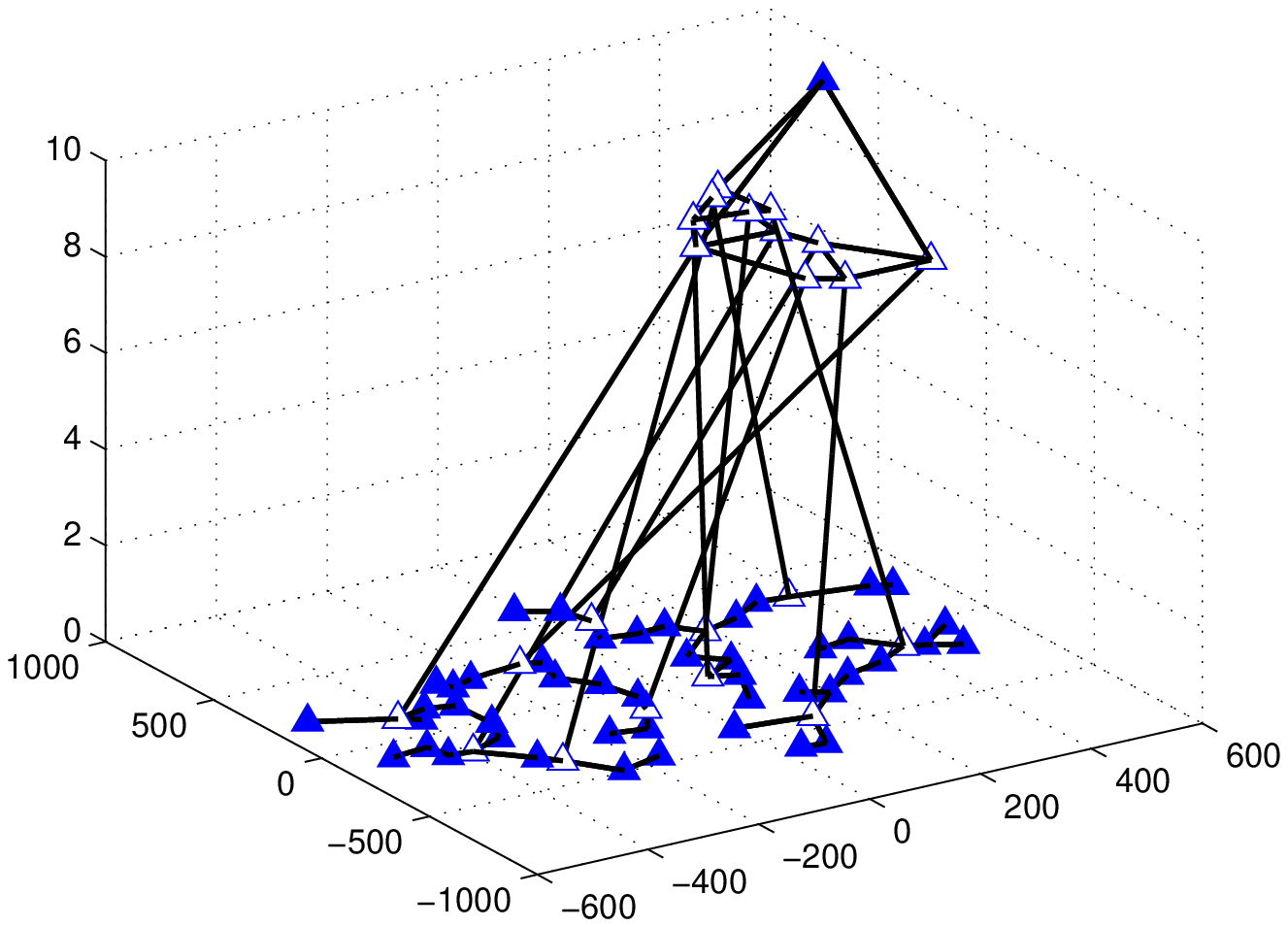}}
}}
\end{center}
\caption{The considered network consists of 57 BSs and 11 routers.
Fig. \ref{FigTopology} (a) plots the locations and the connectivity
of all the BSs. Here, the solid triangles denote BSs, which only
connect to other BSs, and the hollow triangles denote the gateway BSs
that are connected to routers and other BSs.
Fig. \ref{FigTopology} (b) plots the connections between BSs and
routers, which are displayed in the upper part of the graph.
}\label{FigTopology}
\end{figure}

In this section, we report some numerical results on the performance of the proposed algorithms as applied to a network with 57 BSs and 11 network routers. We have tested both the
the efficacy and the efficiency of the proposed algorithms. The topology and the connectivity of this network are shown in Fig. \ref{FigTopology}. For the backhaul links of this network, a fixed capacity is assumed, and is same in both directions. These link capacities are given as follows:
\begin{itemize}
\item links between routers and those between gateway BSs and the routers: 1 (Gnats/s);
\item 1-hop to the gateways: 100 (Mnats/s);
\item 2-hop to the gateways: [10,50] (Mnats/s);
\item 3-hop to the gateways: [2,5] (Mnats/s);
\item More than 4-hop to the gateways: 0 (nats/s).
\end{itemize}

The number of subchannels is
$K=3$ and each subchannel has $1$ MHz bandwidth. The power
budget for each BS is chosen equally by $\bar p=p_{s}$,
$\forall\;s\in\mathcal{B}$, and
$\sigma_{d}^{2}=1,~\forall\;d\in\mathcal{U}$. The wireless links
follow the Rayleigh distribution with $CN(0,(200/{\rm dist})^3)$, where
${\rm dist}$ is the distance between BS and the corresponding user. The
source (destination) node of each commodity is randomly selected
from network routers (mobile users), and all simulation results are
averaged over $100$ randomly selected end-to-end commodity pairs.
Below we refer to one round of the N-MaxMin iteration as an {\it
outer iteration}, and one round of Algorithm 1 for solving
$(\mathbf{r},\mathbf{p})$ as an {\it inner iteration}.

\begin{figure}
\begin{center}
{\resizebox{0.45\textwidth}{!}{\includegraphics{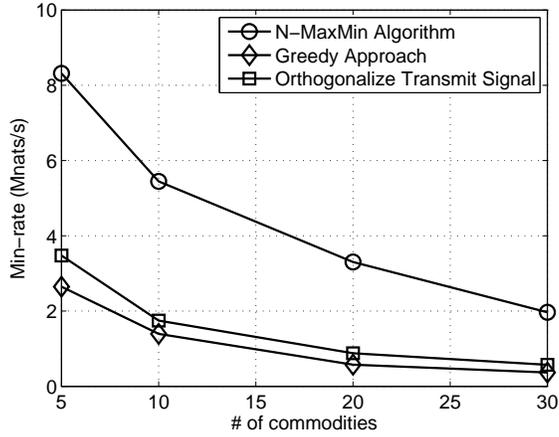}}
}
\end{center}\vspace{-0.2cm}
\caption{The minimum rate achieved by N-MaxMin algorithm and the two
heuristic algorithms for different number of commodities. We have
$\bar p=20$dB.}\label{FigCompare}\vspace{-0.3cm}
\end{figure}

In the first  experiment, we assume that all mobile users can be
served by BSs within $300$ meters and are interfered by
all BSs. For this problem, the parameters of N-MaxMin algorithm are
set to be $\rho_{1}=0.1$ and $\rho_{2}=0.001$; the termination
criterion is
\begin{align}
&\frac{(r^{(t+1)}+\hat r^{(t+1)})-(r^{(t)}+\hat
r^{(t)})}{r^{(t)}+\hat r^{(t)}}<10^{-3}\nonumber\\
&\max\{\|\mathbf{C}{\bf r}_{\rm stack}^{(t)}-\hat{\bf
r}_{\rm stack}^{(t)}\|_{\infty},\|(\mathbf{D}{\bf
p}_{\rm stack}^{(t)})^{2}-(\hat{\bf
p}_{\rm stack}^{(t)})^{2}\|_{\infty}\}\}<5\times 10^{-4}
\end{align}
where $(\cdot)^{2}$ represents elementwise square operation.

For comparison purpose, the following two heuristic algorithms are
considered.
\begin{itemize}
\item {\bf Heuristic 1 (greedy approach)}:\\
We assume that each mobile user is
served by a single BS on a specific frequency tone. For each user, we pick the BS and channel pair that
has the strongest channel as its serving BS and channel.
After BS-user association is determined, each
BS uniformly allocates its power budget to the available frequency tones as well as to
the served users on each tone.
With the obtained power allocation and BS-user
association, the capacity of all wireless links are available and
fixed, so the minimum rate of all commodities can be maximized by
solving a  multi-commodity routing problem (which is essentially
problem \eqref{MainProb} with only backhaul links and network routers).

\item {\bf Heuristic 2 (orthogonal wireless transmission)}:\\
For the second heuristic algorithm, each BS uniformly allocates its
power budget to each frequency tone. To obtain a tractable problem
formulation, we further assume that each active wireless link is
interference free. By doing this each wireless link rate constraints
of problem \eqref{MainProb} now becomes convex. To impose this
interference free constraint, additional variables
$\beta_{l}\in\{0,1\},~\forall\; l\in\mathcal{L}^{wl}$ are introduced,
where $\beta_{l}=1$ if wireless link $l$ is active, otherwise
$\beta_{l}=0$. In this way, there is no interference on wireless link
$l$ if $\sum_{n\in I(l)}\beta_{n}=1$. To summarize, we solve the following optimization
problem:
\begin{align}\label{HeuristicRouting}
\max~&r\notag\\
{\rm s.
t.}~&r_{m}\geq r,~r_{l}(m)\geq
0,~m=1\sim M,~\forall\; l\in\mathcal{L}\notag\\
&\sum_{m=1}^{M}r_{l}(m)\leq
\beta_{l}\log\left(1+\frac{|h_{ds}^{k}|^{2}\bar
p_{s}/K}{\sigma_{d}^{2}}\right),\forall \; l=(s,d,k)\in\mathcal{L}^{wl}\notag\\
&\sum_{n\in I(l)}\beta_{n}=1, ~\beta_{l}\in\{0,1\},~\forall\;
l,n\in\mathcal{L}^{wl},\\
&\eqref{CapacityWired} \mbox{~and~} \eqref{ConservationConst}.\notag
\end{align}
Since the integer constraints on
$\{\beta_{l}\mid\forall~l\in\mathcal{L}^{wl}\}$ are also
intractable, we relax it to $\beta_{l}=[0,1]$. In this way the problem
becomes a large-scale LP, whose solution represents an upper
bound value of problem \eqref{HeuristicRouting}.
\end{itemize}

In Fig.~\ref{FigCompare}, we show the minimum rate performance of
different algorithms when $\bar p=20$dB and $M=5\sim30$. We observe
that the minimum rate achieved by the N-MaxMin algorithm is more
than twice of those achieved by the heuristic algorithms. 

\begin{figure}
\begin{center}
{ \subfigure[]
[]{\resizebox{.45\textwidth}{!}{\includegraphics{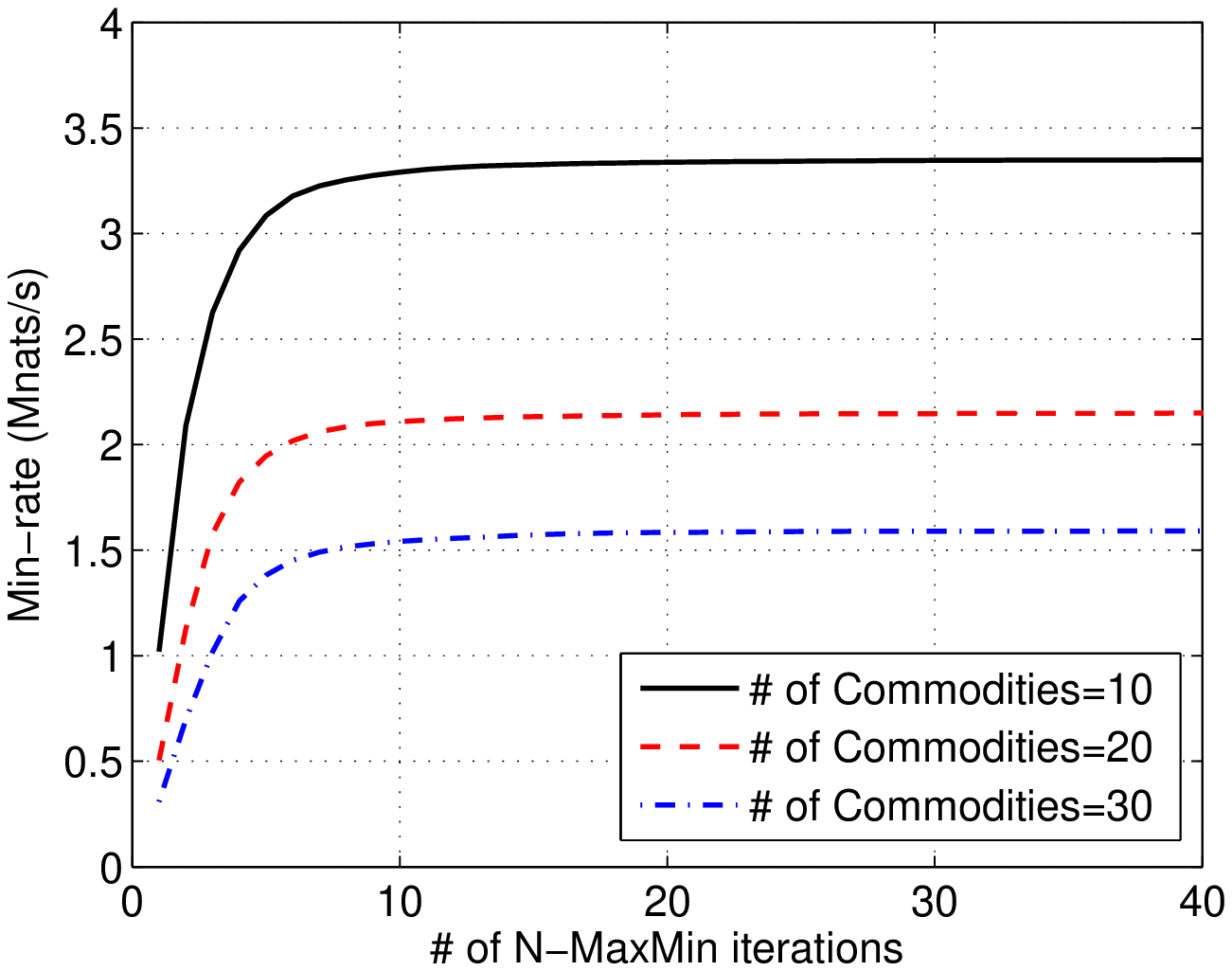}}}}
\hspace{1pc} { \subfigure[][
]{\resizebox{.45\textwidth}{!}{\includegraphics{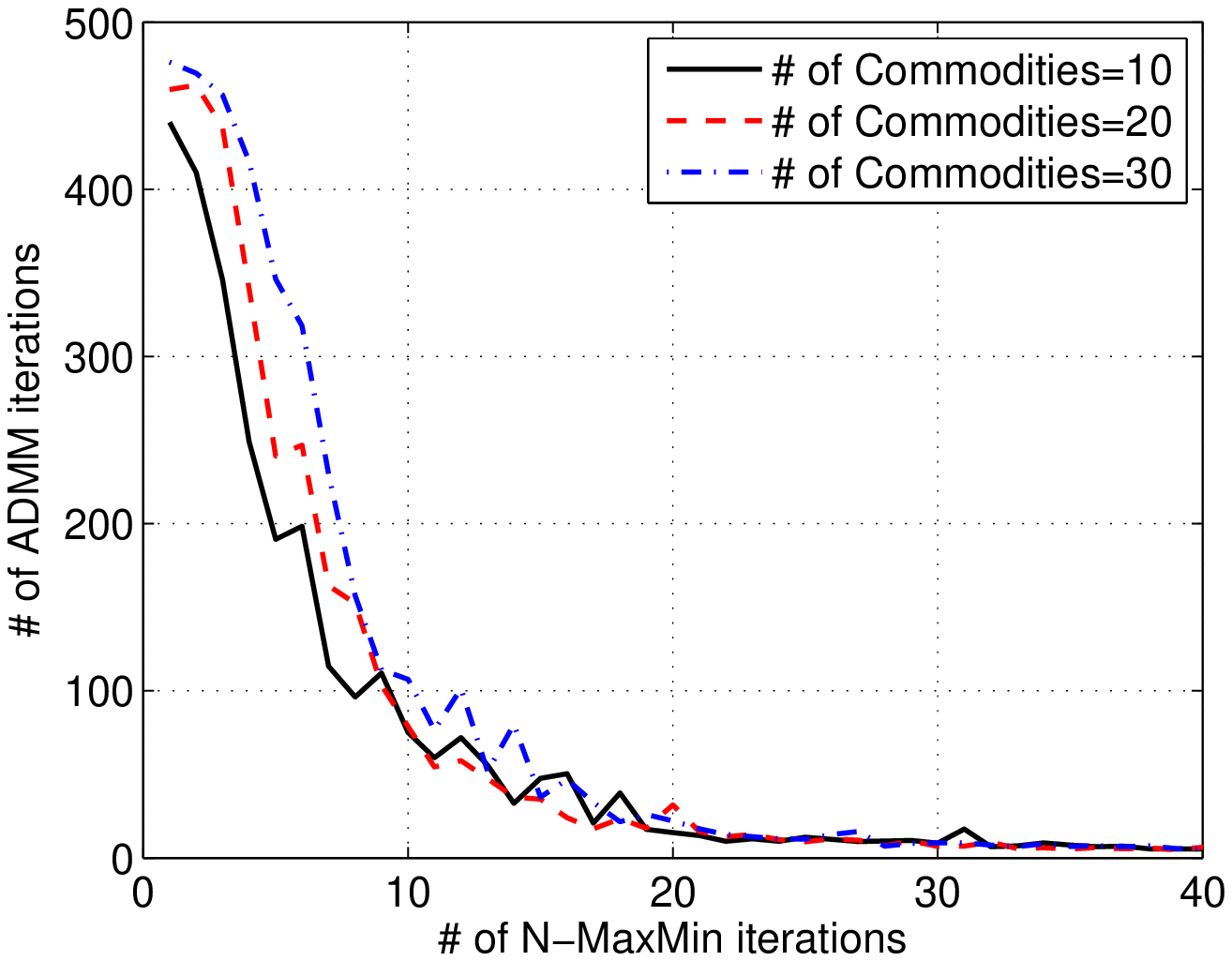}}
}}\hspace{1pc}{ \subfigure[]
[]{\resizebox{.45\textwidth}{!}{\includegraphics{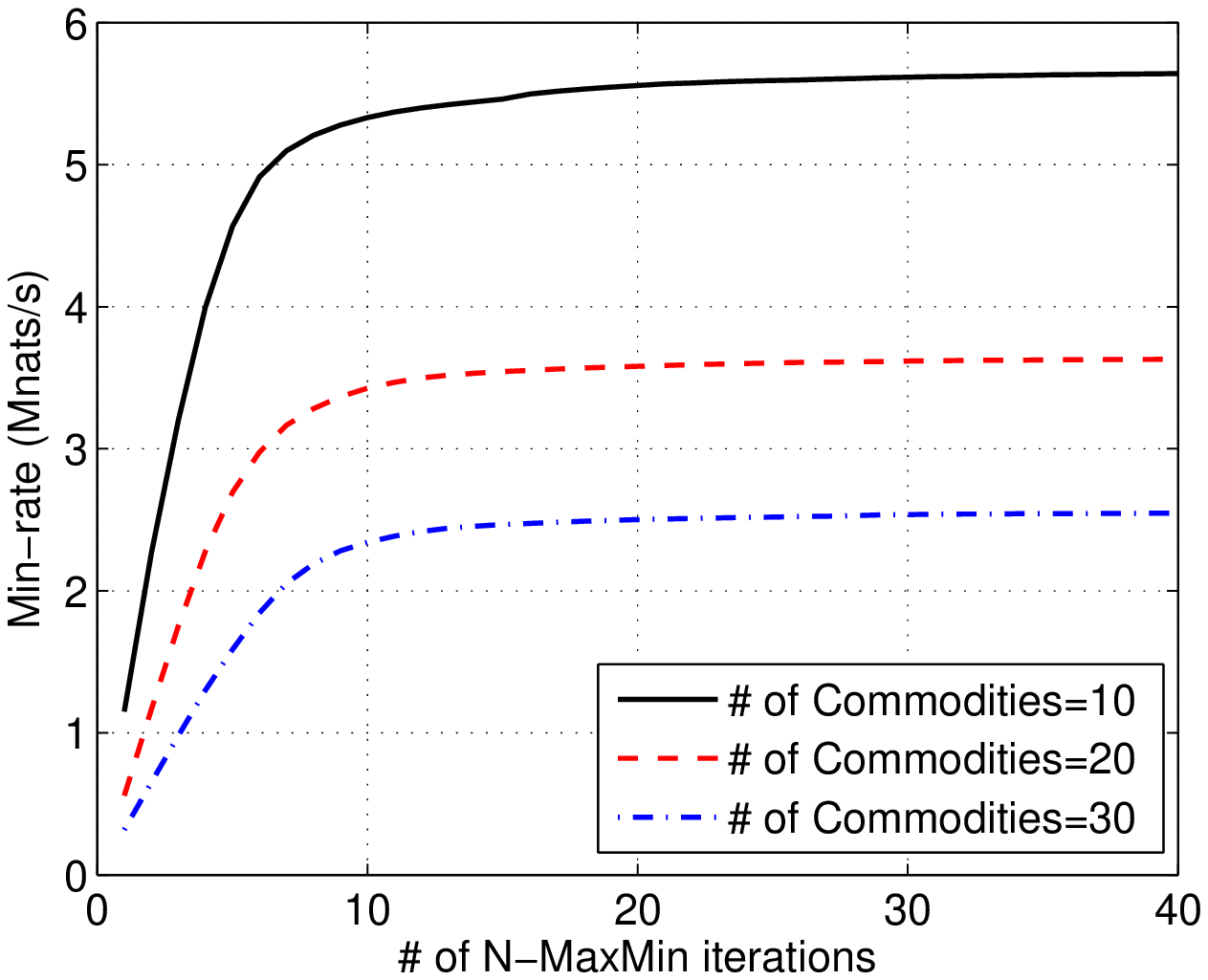}}}}
\hspace{1pc} { \subfigure[][
]{\resizebox{.45\textwidth}{!}{\includegraphics{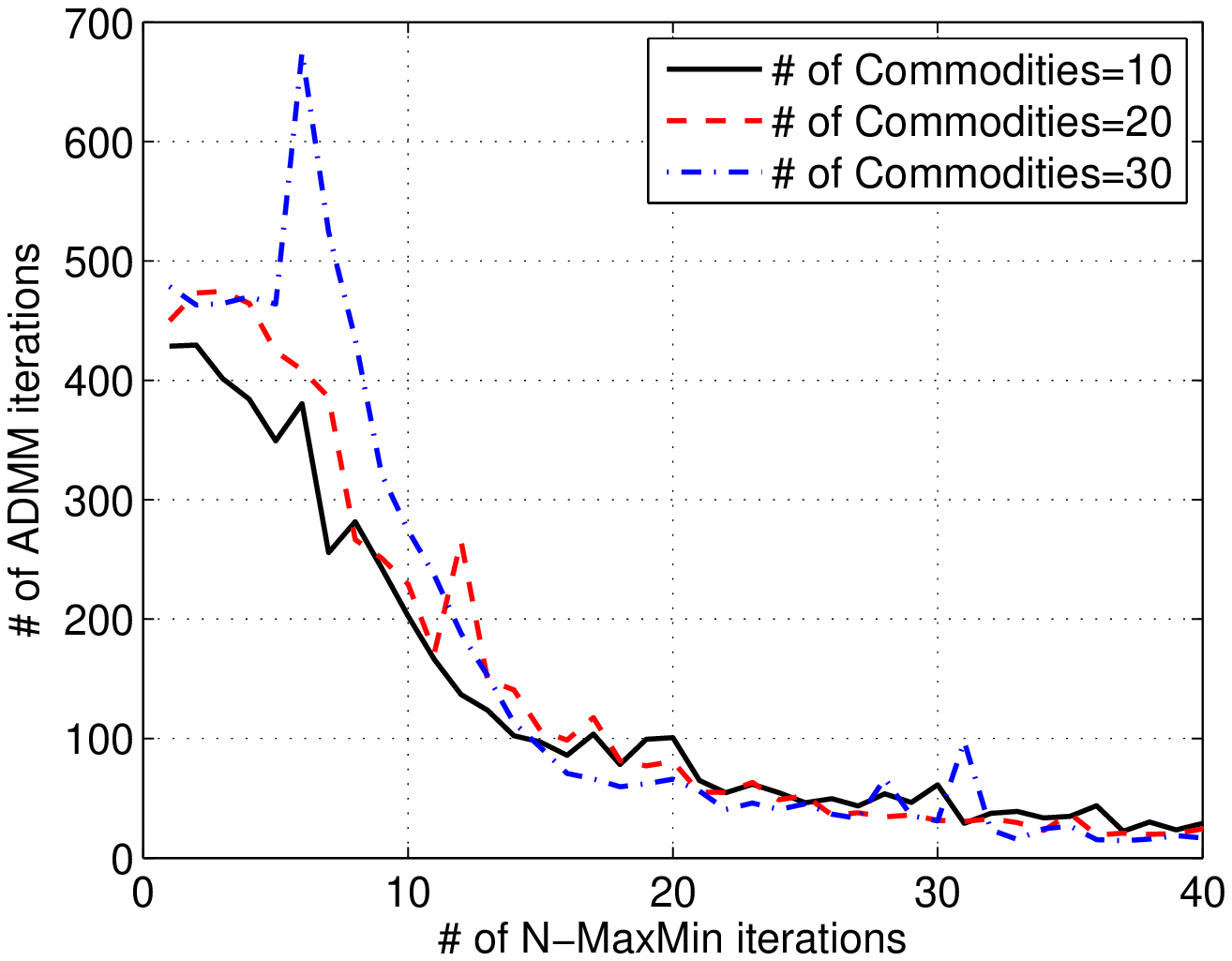}}
}} \hspace{1pc}
\end{center}\vspace{-0.3cm}
\caption{\footnotesize The minimum rate performance and the required
number of iterations for the proposed N-MaxMin algorithm. In
[(a)(b)] $\bar p=10$dB and in [(c)(d)] $\bar p=20$dB. In [(a)(c)],
the obtained minimum rate versus the iterations of N-MaxMin is
plotted. In [(b)(d)], the required number of inner ADMM iterations
is plotted against the iteration for the outer N-MaxMin
algorithm.}\label{FigRateIte}\vspace{-0.3cm}
\end{figure}

In the second set of numerical experiments, we evaluate the proposed
N-MaxMin algorithm using different number of commodity pairs and
different power budgets at the BSs. Here we use the same settings as
in the previous experiment, except that all mobile users are
interfered by the BSs within a distance of $800$ meters, and that we set
$\rho_{2}=0.005$ (resp. $\rho_2=0.001$)  when $\bar p=10$ dB (resp.~$\bar p=20$ dB).
The minimum rate performance for the N-MaxMin
algorithm and the required number of inner iterations are plotted in
Fig. \ref{FigRateIte}. Due to the fact that the obtained $\{{\bf
r},{\bf p}\}$ is far from the stationary solution in the first few
outer iterations, there is no need to complete Algorithm 1 at the
very beginning. Hence, we limit the number of inner iterations to be
no more than $500$ for the first $5$ outer iterations. After the
early termination of the inner Algorithm 1, we use the obtained
${\bf p}$ to update ${\bf u}$ and ${\bf w}$ by \eqref{UpdateU} and
\eqref{UpdateW}, respectively.

In Fig. \ref{FigRateIte}(a)--(b), we see that when $\bar p=10$ dB,
the minimum rate converges at about the $10$th outer iteration when
the number of commodities is up to $30$, while less than $500$ inner
iterations are needed per outer iteration. Moreover, after the
$10$th outer iteration, the number of inner ADMM iterations reaches
below $100$. In Fig. \ref{FigRateIte}(c)--(d), the case with $\bar
p=20$dB is considered. Clearly the required number of outer
iterations is slightly more than that in the case of $\bar p=10$dB,
since the objective value and the feasible set are both larger.
However, in all cases the algorithm still converges fairly quickly.

\begin{figure}[t]
\begin{center}
{ \subfigure[]
[]{\resizebox{.45\textwidth}{!}{\includegraphics{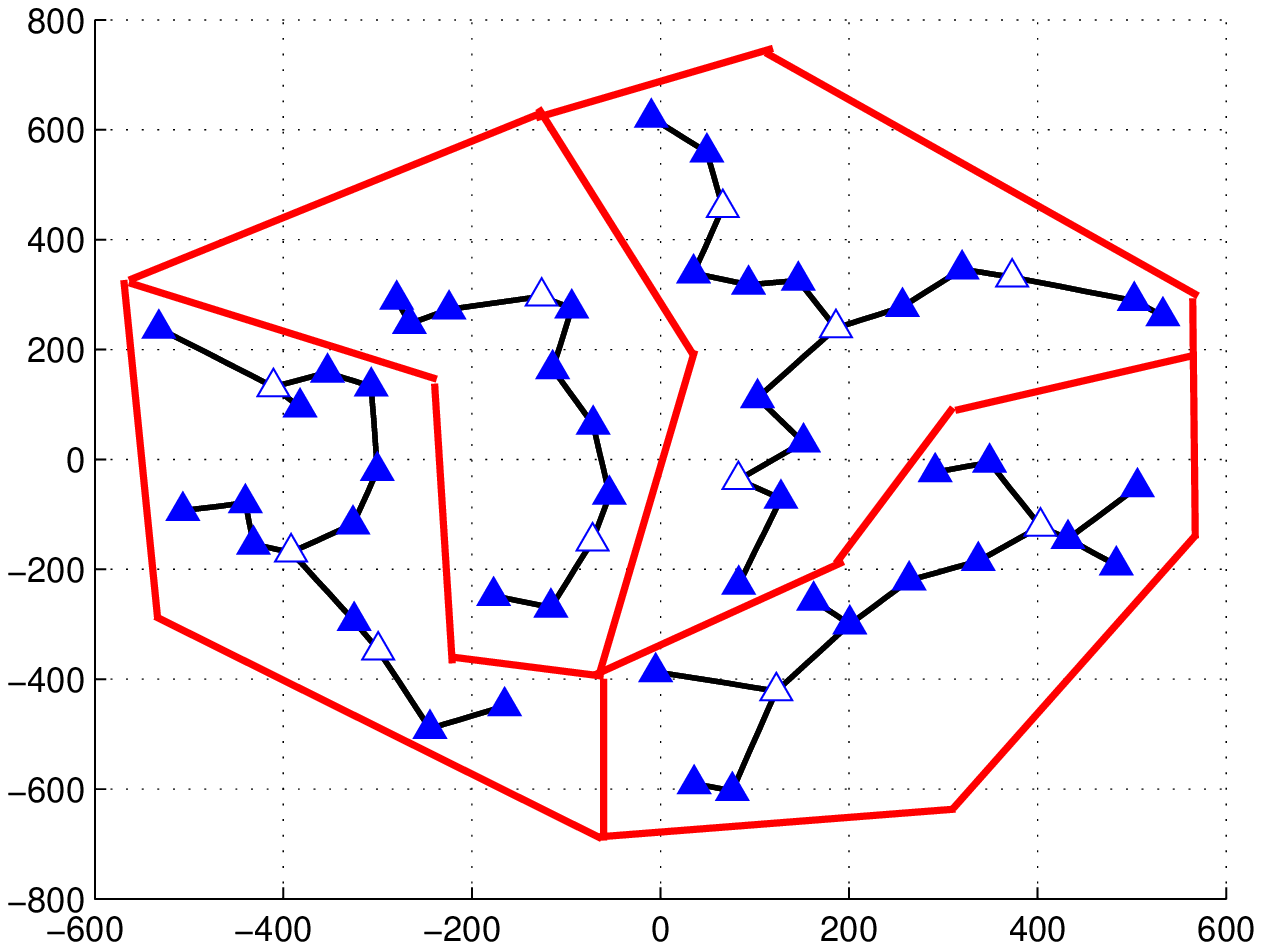}}}}
\hspace{1pc} { \subfigure[][
]{\resizebox{.45\textwidth}{!}{\includegraphics{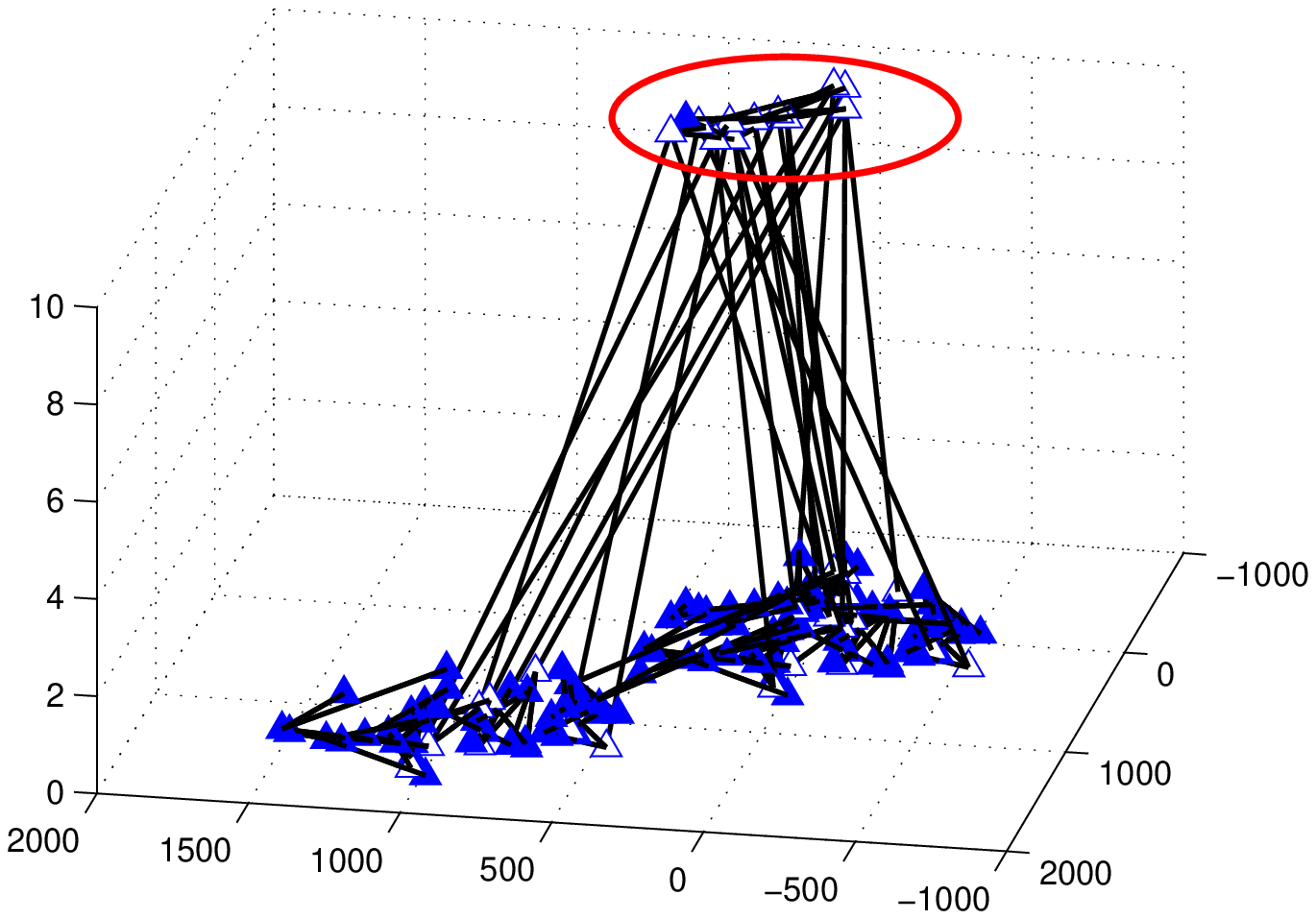}}
}}
\end{center}
\caption{The considered network consists of 114 BSs and 12 routers. Each computation core is responsible for
one group of nodes shown in the figure. Fig.
\ref{FigCore} (a) plots the locations and the connectivity of a single
basic BS set (consists of 57 BSs). The solid triangles denote the BSs, which only
connect to other BSs, and the hollow triangles denote the BSs
serving the gateways that are connected to routers and other BSs.
Fig. \ref{FigCore} (b) displays the connections between the BSs and routers. }\label{FigCore}
\end{figure}

In the last set of numerical experiments, we demonstrate how parallel implementation
can speed up Algorithm 1 considerably. To illustrate the benefit of parallelization,
we consider a larger network (see Fig.\
\ref{FigCore} (b)) which is derived by merging two identical
networks shown in Fig.\ \ref{FigTopology} (a). The new network
consists of 126 nodes (12 network routers and 114 BSs).

%
For simplicity, we removed all the wireless links, so constraints \eqref{CapacityWireless} and
\eqref{PowerConstraint} of problem \eqref{MainProb} are absent.
This reduces problem \eqref{MainProb} to a network flow problem (a very large linear program).

We implement Algorithm 1 using the Open MPI package, and compare its
efficiency with the commercial LP solver, Gurobi \cite{gurobi}. For
the Open MPI implementation, we use 4 computation cores for each
basic BS set as illustrated in Fig.\ \ref{FigCore} (a), and use 1
additional computation core for all the network routers shown in
Fig.\ \ref{FigCore} (b). Since we have two identical subnetworks
(connected by a common set of routers), we have in total 9
computation cores. We choose $\rho_{1}=0.01$ and let the BSs serve
as the destination nodes for commodities. Table
\ref{TableTimeCompare} compares the computation time required for
different implementation of Algorithm 1 and that of Gurobi. We
observe that parallel implementation of Algorithm 1 leads to more
than 5 fold improvement in computation time computed on SunFire
X4600 server with AMD Opteron 8356 2.3GHz CPUs. We also note that
when the problem size increases, the performance of Gurobi becomes
worse than that achieved by the parallel implementation of Algorithm
1. Thus, the proposed algorithm (implemented in parallel) appears to
scale nicely to large problem sizes.


%
%

\begin{table}[th]
\begin{center}
\begin{tabular}{|c|c|c|c|c|c|}
\hline $\begin{array}{c}\mbox{\# of} \\
\mbox{Commodities} \end{array}$ & 50 & 100 & 200 & 300 \\
\hline
$\begin{array}{c}\mbox{Time (s)} \\
\mbox{(Sequential)} \end{array}$ & 1.04 & 2.03 & 4.73 & 8.53 \\
\hline
$\begin{array}{c}\mbox{Time (s)} \\
\mbox{(Parallel)} \end{array}$ & 0.20 & 0.37 & 0.75 & 1.10 \\
\hline
$\begin{array}{c}\mbox{Time (s)} \\
\mbox{(Gurobi)} \end{array}$ & 0.20 & 0.64 & 1.65 & 2.51 \\
\hline
$\begin{array}{c}\mbox{\# of} \\
\mbox{Variables} \end{array}$ &
1.4$\times10^{4}$ & 2.9$\times10^{4}$ & 5.8$\times10^{4}$ & 8.7$\times10^{4}$\\
\hline
$\begin{array}{c}\mbox{\# of} \\
\mbox{Constraints} \end{array}$ & 2.1$\times10^{4}$ & 4.2$\times10^{4}$ & 8.4$\times10^{4}$ & 1.3$\times10^{5}$ \\
\hline
\end{tabular}
\end{center}
\caption{Comparison of computation time used by different implementations of Algorithm 1 for the routing only problem. The size of the problems solved are specified using a range of metrics (total number of commodities, variables and constraints). }\label{TableTimeCompare}\vspace{-1cm}
\end{table}

\section{Concluding Remarks}\label{Sec:Conclusion}

In this paper, we have considered the joint backhaul traffic engineering and interference management problem for a SD-RAN. In the considered problem, the resources in both the fixed backhaul links and the wireless radio access links are optimized.  
Although the problem is nonconvex, large-scale, and the optimization
variables are coupled in various constraints, our proposed
algorithm is capable of efficiently computing a high-quality solution in a
distributed manner. Key to the efficiency of the proposed algorithm
is the use of the well-known rate-MSE relationship, which helps
transform the original problem into a form that is amendable to
alternating optimization. In each iteration of the algorithm, two
separate subproblems are solved, one admits a closed-form solution,
while the other can be solved efficiently by using the ADMM
algorithm. The proposed algorithm is scalable to large networks since
all its steps can be computed in closed-form independently
and in parallel across all nodes of the network. Simulation results show that the proposed algorithm
significantly outperforms heuristic algorithms in terms of the achieved min-rate. As a future work, we plan to
investigate the use of stochastic WMMSE algorithm \cite{SWMMSE} to reduce the amount of channel state information.

\appendix
\def\thesection{\Alph{section}}
\subsection{Proof of Theorem
\ref{BCDConverge}:}\label{Appendix:WMMSEProof}

This proof follows a similar argument as in
\cite{Razaviyayn13}, so here we only provide the main steps of the
proof. For the following discussion, we denote the KKT solutions of
problem \eqref{MainProb} as $\{{\bf r}^{\star},{\bf
p}^{\star};{\bfdelta}^{\star},{\bftheta}^{\star},{\bfepsilon}^{\star},{\bf
\kappa}^{\star}\}$ where
${\bfdelta}^{\star},{\bftheta}^{\star},{\bfepsilon}^{\star}$, and
${\bfkappa}^{\star}$ respectively denotes the corresponding
Lagrangian dual variables for the nonnegativeness constraints
$\{r\geq 0,~r_{m}\geq r,~r_{l}(m)\geq 0\mid l\in\mathcal{L},m=1\sim
M\},$ as well as
\{\eqref{CapacityWired},~\eqref{CapacityWireless}\},
\eqref{PowerConstraint} and \eqref{ConservationConst}. For problem
\eqref{WMMSEQ}, the KKT solutions are similarly denoted as
$\{\hat{\bf r},\hat{\bf p},\hat{\bf u},\hat{\bf
w};\hat{\bfdelta},\hat{\bftheta},\hat{\bfepsilon},\hat{\bfkappa}\}$,
where $\hat{\bfdelta}$ now is the Lagrangian dual variables for constraints
\eqref{CapacityWired} and \eqref{RateMSEConstraint}.

{\bf Step 1: If $\xb^{\star}\triangleq\{{\bf r}^{\star},{\bf
p}^{\star};{\bfdelta}^{\star},{\bftheta}^{\star},{\bfepsilon}^{\star},{\bfkappa}^{\star}\}$
is an arbitrary KKT solution of problem \eqref{MainProb},
$\{\hat{\bf r},\hat{\bf p},\hat{\bf u},\hat{\bf
w};\hat{\bfdelta},\hat{\bftheta},\hat{\bfepsilon},\hat{\bfkappa}\}$
chosen as
$$\yb^{\star}\triangleq\{{\bf
r}^{\star},{\bf p}^{\star},{\bf u}({\bf p}^{\star}),{\bf w}({\bf
p}^{\star});{\bfdelta}^{\star},{\bftheta}^{\star},{\bfepsilon}^{\star},{\bfkappa}^{\star}\}$$
is also a KKT solution of problem \eqref{WMMSEQ}. The converse
statement is also true. Here ${\bf u}({\bf p^{\star}})$ and ${\bf
w}({\bf p^{\star}})$ are the ${\bf u}$ and ${\bf w}$ obtained by
\eqref{UpdateU} and \eqref{UpdateW} for a given ${\bf p^{\star}}$. }

Since some of the constraints of problem \eqref{MainProb} and
problem \eqref{WMMSEQ} are the same, i.e., \eqref{CapacityWired},
\eqref{PowerConstraint}, and \eqref{ConservationConst}, the
corresponding feasibility and the complementary slackness conditions
of these constraints are of the same form for both problems. Hence,
if $\xb^{\star}$ can satisfy these constraints for problem
\eqref{MainProb}, $\yb^{\star}$ can satisfy those of problem
\eqref{WMMSEQ}. Hence, we should only consider the remaining KKT
conditions given below. For problem \eqref{MainProb}, we have
\begin{subequations}\label{KKTWMMSEP}
\begin{align}
&-2\epsilon_{s}^{\star}p_{ds}^{k\star}+\sum_{n=(s',d',k)\in\bar I(s,d,k)}\theta_{d's'}^{k\star}\nabla_{p_{ds}^{k}}\bar r_{n}({\bf p}^{\star})=0,~\forall\; l=(s,d,k)\in\mathcal{L}^{wl},\\
&\delta^{\star}+\sum_{m=1}^{M}\delta_{m}^{\star}=1,\label{KKTWMMSEP2}\\
&\delta_{m}^{\star}+\kappa^{S(m)\star}(m)-\kappa^{D(m)\star}(m)=0,~m=1\sim M,\label{KKTWMMSEP3}\\
&\delta_{l}^{\star}(m)-\theta_{l}^{\star}+\sum_{v:l\in{\rm
In}(v)}\kappa^{v\star}(m)-\sum_{v:l\in{\rm
Out}(v)}\kappa^{v\star}(m)=0,~\forall\; v\in\mathcal{V},~m=1\sim M,\label{KKTWMMSEP4}\\
&0\leq \theta_{l}^{\star}\ \bot\ \bar r_{l}({\bf
p}^{\star})-\sum_{m=1}^{M}r_{l}^{\star}(m)\geq 0,~\forall\;
l\in\mathcal{L}^{wl}.\label{KKTWMMSEP5}
\end{align}
\end{subequations}
For problem \eqref{WMMSEQ}, we have
\begin{subequations}\label{KKTWMMSEQ}
\begin{align}
&-2\hat\epsilon_{s}\hat p_{ds}^{k}+\sum_{n=(s',d',k)\in
I(s,d,k)}\hat\theta_{d's'}^{k}\nabla_{p_{ds}^{k}}E_{n}(\hat u_{n},\hat w_{n}, \hat{\bf p})=0,\label{KKTWMMSEQ1}\\
&\hat\theta_{ds}^{k}\nabla_{u_{l}}E_{l}(\hat u_{l},\hat w_{l}, \hat{\bf p})=0,\label{KKTWMMSEQ2}\\
&\hat\theta_{ds}^{k}\nabla_{w_{l}}E_{l}(\hat u_{l},\hat w_{l}, \hat{\bf p})=0,~\forall\; l=(s,d,k)\in\mathcal{L}^{wl},\label{KKTWMMSEQ3}\\
&\hat\delta+\sum_{m=1}^{M}\hat\delta_{m}=1,\label{KKTWMMSEQ4}\\
&\hat\delta_{m}+\hat\kappa^{S(m)}(m)-\hat\kappa^{D(m)}(m)=0,~m=1\sim M,\label{KKTWMMSEQ5}\\
&\hat\delta_{l}(m)-\hat\theta_{l}+\sum_{v:l\in{\rm
In}(v)}\hat\kappa^{v}(m)-\sum_{v:l\in{\rm
Out}(v)}\hat\kappa^{v}(m)=0,~\forall\; v\in\mathcal{V},~m=1\sim M,\label{KKTWMMSEQ6}\\
&0\leq \hat\theta_{l}\ \bot\ E_{l}(\hat u_{l},\hat w_{l}, \hat{\bf
p})-\sum_{m=1}^{M}\hat r_{l}(m)\geq 0,~\forall\;
l\in\mathcal{L}^{wl}.\label{KKTWMMSEQ7}
\end{align}
\end{subequations}

Obviously, by comparing \eqref{KKTWMMSEP2}$\sim$\eqref{KKTWMMSEP4}
and \eqref{KKTWMMSEQ4}$\sim$\eqref{KKTWMMSEQ6}, we can conclude that
$\yb^{\star}$ can satisfy
\eqref{KKTWMMSEQ4}$\sim$\eqref{KKTWMMSEQ6}. For \eqref{KKTWMMSEQ2}
and \eqref{KKTWMMSEQ3}, by the optimality of \eqref{UpdateU} and
\eqref{UpdateW}, they are also true for $\yb^{\star}$. Moreover, since $\bar r_{l}({\bf
p}^{\star})=E_{l}(u_{l}({\bf p}^{\star}),w_{l}({\bf p}^{\star}),
{\bf p}^{\star})$, it follows from Lemma \ref{Rate-MSE} that
\eqref{KKTWMMSEQ7} can be satisfied with $\yb^{\star}$.

For the last KKT condition of problem \eqref{WMMSEQ}, i.e.,
\eqref{KKTWMMSEQ1}, let us first split the Lagrange multiplier
$\bftheta^{\star}$ into two subsets
\begin{align*}
\mathcal{A}&\triangleq\{l\mid \theta_{l}^{\star}>0,~\forall\;
l\in\mathcal{L}\},\quad\mathcal{\bar A}\triangleq\{l\mid
\theta_{l}^{\star}=0,~\forall\;l\in\mathcal{L}\}.
\end{align*}
Then by the same argument as Proposition 1 in
\cite{Razaviyayn13}, \eqref{KKTWMMSEQ1} is also satisfied by
$\yb^{\star}$. The reverse statement of step 1 can be argued
similarly.

{\bf Step 2: Every global optimal solution of problem \eqref{MainProb} corresponds to a global optimal solution of problem \eqref{WMMSEQ}, and they achieve the same objective value.}

 To show this step, we recall that the network is connected and the link capacities are positive. It follows that the optimal value $r^{\star}$ must be strictly greater than 0. Hence, the corresponding Lagrangian dual variable $\delta^{\star}$ is always $0$ by the complementarity condition, and the KKT condition \eqref{KKTWMMSEP2} becomes $\sum_{m=1}^{M}\delta_{m}^{\star}=1$. The argument is the same for $\hat\delta$, so $\sum_{m=1}^{M}\hat\delta_{m}=1$. With this fact, we can use the proof of Proposition 3 in \cite{Razaviyayn13} to show the desired result.

{\bf Step 3: The proposed alternating optimization method can converge to the KKT solutions of problem \eqref{MainProb}.}

Given the results of previous two steps and by Theorem 2 of \cite{Razaviyayn13}, the final convergence result is proved. \hfill \ensuremath{\Box}

\subsection{Derivation of Updating Steps of Algorithm
1}\label{Sec:StepByStep}
In this section, we go over Algorithm 1 step by step and explain
each of its update procedure. For notational simplicity, we
ignore the superscript indices.

\subsubsection{\bf Solving Step 3 for Algorithm 1}

In this step, problem
\eqref{WMMSEQFirst} is solved to update $\{{\bf r},\hat{\bf p}\}$.
This problem can be further decomposed over the variables $\{r,r_{m},r_{l}(m)\mid
m=1\sim M,\;\forall\; l\in\mathcal{L}^{w}\}$ and $\hat{\bf
p}\cup \{r_{l}(m)\mid m=1\sim M,\;\forall\; l\in\mathcal{L}^{wl}\}$.

The first subblock only has to do with the wired links. A closer look at Step 3 of Algorithm 1 reveals that its optimization problem can be solved via two completely independent subproblems, one for variables $\{r,r_{m}\mid m=1\sim M\}$ and the other for $\{r_{l}(m)\mid m=1\sim M,\forall \;l\in\mathcal{L}^w\}$. In the following we consider the two problems separately.

{\bf (i) Subproblem for $\{r,r_{m}\mid m=1\sim M\}$}:
This step updates the current minimum flow rate among all commodities, and it can be mathematically expressed as{\small
\begin{align}\label{UpdateFirstSub1}
\max ~&\frac{r}{2}-\frac{\rho_{1}}{2}\left(\hat
 r-r-\frac{\delta}{\rho_{1}}\right)^{2}-\frac{\rho_{1}}{2}\sum_{m=1}^{M}\left[\left(\hat
 r_{m}^{S(m)}-r_{m}-\frac{\delta_{m}^{S(m)}}{\rho_{1}}\right)^{2}+\left(\hat
 r_{m}^{D(m)}-r_{m}-\frac{\delta_{m}^{D(m)}}{\rho_{1}}\right)^{2}\right]\notag\\
{\rm s. t.}~&r_{m}\geq r,~m=1\sim M,~r\geq 0.
\end{align}}
When $r$ is fixed, the optimal $\{r_{m}^{\star}\}_{m=1}^M$ of problem \eqref{UpdateFirstSub1} can be obtained by the first-order optimality condition as follows
\begin{align}
&r_{m}^{\star}=\frac{1}{2}\left(\hat r_{m}^{S(m)}+\hat r_{m}^{D(m)}-\frac{\delta_{m}^{S(m)}+\delta_{m}^{D(m)}}{\rho_{1}}+\frac{\lambda_{m}^{\star}}{\rho_{1}}\right),~m=1\sim M,
\end{align}
where $\{\lambda_{m}^{\star}\geq 0\}$ are the optimal Lagrange dual variables for constraints $\{r_{m}\geq r\}$. Due to the complementarity condition, and the fact that $r_{m}^{\star}$ is an increasing function of $\lambda_{m}^{\star}$, it follows that $\lambda_{m}^{\star}>0$ only if the equality holds for $r_{m}^{\star}\geq r$. Thus, we can conclude
\begin{align}\label{UpdateFirstSub1Sol}
r_{m}^{\star}=\max\left\{r,\frac{1}{2}\left(\hat r_{m}^{S(m)}+\hat r_{m}^{D(m)}-\frac{\delta_{m}^{S(m)}+\delta_{m}^{D(m)}}{\rho_{1}}\right)\right\}.
\end{align}
After plugging the obtained $r_{m}^{\star}$ back to the objective function of problem \eqref{UpdateFirstSub1}, the gradient of the objective function with respect to $r$ is given by {\small
\begin{align}\label{Derivativer}
\frac{1}{2}+\rho_{1}\left(\hat
 r-\frac{\delta}{\rho_{1}}\right)
-\rho_{1}\left\{r+2\hspace{-.4cm}\sum_{m:\frac{1}{2}\left(\hat r_{m}^{S(m)}+\hat r_{m}^{D(m)}-\frac{\delta_{m}^{S(m)}+\delta_{m}^{D(m)}}{\rho_{1}}\right)\leq r}\hspace{-.1cm}\left[r-\frac{1}{2}
\left(\hat
 r_{m}^{S(m)}+\hat
 r_{m}^{D(m)}-\frac{\delta_{m}^{S(m)}+\delta_{m}^{D(m)}}{\rho_{1}}\right)\right]\right\}.
\end{align}}
Notice that the obtained derivative is a decreasing function
for $r\geq 0$. Thus, the optimal $r^{\star}=0$ if
\eqref{Derivativer} is no more than 0 with $r=0$. Otherwise,
$r^{\star}$ can be obtained through bisection procedure over $r\geq
0$ such that \eqref{Derivativer} equals 0.

{\bf (ii) Subproblem for $\{r_{l}(m)\mid m=1\sim M, \; \forall
\;l\in\mathcal{L}^w\}$}: It turns out that for this subset of
variables, the corresponding updating procedure can be performed independently
{\it over each link}. For each link $l=(s,d)\in\mathcal{L}^w$, the
following optimization problem is solved
\begin{align}\label{UpdateFirstSub2}
\min~&\sum_{m=1}^{M}\left[\left(\hat
r_{l}^{s}(m)-r_{l}(m)-\frac{\delta_{l}^{s}(m)}{\rho_{1}}\right)^{2}+\left(\hat r_{l}^{d}(m)-r_{l}(m)-\frac{\delta_{l}^{d}(m)}{\rho_{1}}\right)^{2}\right]\notag\\
{\rm s. t.}~&\sum_{m=1}^{M}r_{l}(m)\leq C_{l},~r_{l}(m)\geq 0,~m=1\sim M.
\end{align}
The optimal solution $r_{l}^{\star}(m),~m=1\sim M$, of problem \eqref{UpdateFirstSub2} can be obtained by the first-order optimality condition
\begin{align}\label{UpdateFirstSub2Sol}
r_{l}(m)^{\star}=\frac{1}{2}\left(\hat r_{l}^{s}(m)+\hat r_{l}^{d}(m)-\frac{\delta_{l}^{s}(m)+\delta_{l}^{d}(m)}{\rho_{1}}-\frac{\lambda_{l}^{\star}}{2}\right)^{+}\geq 0,~m=1\sim M
\end{align}
where $\lambda_{l}^{\star}$ is the optimal Lagrange dual variable of the capacity constraint on link $l$. Using the complementarity condition and the fact that the left hand side of the capacity constraint is a decreasing function of $\lambda_{l}^{\star}$, it follows that $\lambda_{l}^{\star}=0$ is true only if
$$\sum_{m=1}^{M}\frac{1}{2}\left(\hat r_{l}^{s}(m)+\hat r_{l}^{d}(m)-\frac{\delta_{l}^{s}(m)+\delta_{l}^{d}(m)}{\rho}\right)^{+}\leq C_{l}.$$
Otherwise, $\lambda_{l}^{\star}$ should be chosen such that the capacity constraint is active, and this $\lambda_{l}^{\star}$ can be obtained by bisection procedure over $\lambda_{l}^{\star}\geq 0$.

{\bf (iii) Subproblem for $\hat{\bf p}\cup \{r_{l}(m)\mid m=1\sim
M,\;\forall\; l\in\mathcal{L}^{wl}\}$}: The rest of variables are related only to the wireless links, and they are in
fact decoupled across the wireless links. To be more specific, the problem
for the wireless link $l=(s,d,k)\in\mathcal{L}^{wl}$ is shown below
\begin{align}\label{WMMSEQFirstSub2}
\min ~&\frac{\rho_{1}}{2}\sum_{m=1}^{M}\left[\left(\hat
r_{l}^{s}(m)-r_{l}(m)-\frac{\delta_{l}^{s}(m)}{\rho_{1}}\right)^{2}+\left(\hat
r_{l}^{d}(m)-r_{l}(m)-\frac{\delta_{l}^{d}(m)}{\rho_{1}}\right)^{2}\right]\notag\\
&+\frac{\rho_{2}}{2}\sum_{n=(s',d',k)\in I(l)}\left(p_{d's'}^{k}-p_{ds,d's'}^{k}-\frac{\theta_{ln}}{\rho_{2}}\right)^{2}\notag\\
{\rm s. t.}~&r_{l}(m)\geq 0,~m=1\sim M\\
&\sum_{m=1}^{M}r_{l}(m)\leq c_{1,l}+c_{2,l}p_{ds,ds}^{k}-\sum_{n=(s',d',k)\in I(l)}c_{3,ln}|p_{ds,d's'}^{k}|^{2}\notag.
\end{align}
The optimal solution of this problem,
$\{r_{l}^{\star}(m),p_{ds,d's'}^{k\star}\mid m=1\sim M,(s',d',k)\in
I(l)\}$, can be obtained by the first-order conditions below
\begin{subequations}\label{WMMSEQFirstSub2Sol}
\begin{align}
r_{l}^{\star}(m)&=\frac{1}{2}\left(\hat r_{l}^{s}(m)+\hat
r_{l}^{d}(m)-\frac{\delta_{l}^{s}(m)+\delta_{l}^{d}(m)+\lambda_{l}^{\star}}{\rho_{1}}\right)^{+},~m=1\sim M,\\
p_{ds,ds}^{k\star}&=\frac{\rho_{2}(p_{ds}^{k}-\frac{\theta_{ll}}{\rho_{2}})+\lambda_{l}^{\star} c_{2,l}}{\rho_{2}+2\lambda_{l}^{\star} c_{3,ll}},\\
p_{ds,d's'}^{k\star}&=\frac{\rho_{2}(p_{d's'}^{k}-\frac{\theta_{ln}}{\rho_{2}})}{\rho_{2}+2\lambda_{l}^{\star}
c_{3,ln}},~\forall\; n=(s',d',k)\in I(l),~n\neq l.
\end{align}
\end{subequations}
where $\lambda_{l}^{\star}$ is the optimal Lagrange dual variable
for the rate-MSE constraint.

After plugging the obtained optimal solutions \eqref{WMMSEQFirstSub2Sol} into the rate-MSE constraint of problem \eqref{WMMSEQFirstSub2}, it can be observed that the left hand side of the constraint, $\sum_{m=1}^{M}r_{l}^{\star}(m)$, is a decreasing function of $\lambda_{l}^{\star}$. Furthermore, taking the gradient of the right hand side of the rate-MSE constraint with respect to $\lambda_{l}^{\star}$ gives
\begin{align}
&\frac{\partial (c_{1,l}+c_{2,l}p_{ds,ds}^{k\star}-\sum_{n=(s',d',k)\in I(l)}c_{3,ln}|p_{ds,d's'}^{k\star}|^{2})}{\partial \lambda_{l}^{\star}}\notag\\
=&\frac{1}{\rho_{2}}\left[\frac{\left(c_{2,l}-2c_{3,ll}(p_{ds}^{k\star}-\frac{\theta_{ll}}{\rho_{2}})\right)^{2}}{\left(1+\frac{2}{\rho_{2}}\lambda_{l}^{\star}
 c_{3,ll}\right)^{3}}+\sum_{n=(s',d',k)\in I(l)\setminus\{l\}}\frac{\left(2c_{3,ln}(p_{ds}^{k}-\frac{\theta_{ln}}{\rho_{2}})\right)^{2}}{\left(1+\frac{2}{\rho_{2}}\lambda_{l}^{\star} c_{3,ln}\right)^{3}}\right]\geq 0,
\end{align}
where the nonnegativity is due to the fact that $c_{3,ln}\geq 0$,
$\forall\; n$. Hence, the right hand side of the rate-MSE constraint
is an increasing function of $\lambda_{l}^{\star}\geq 0$. By the
complementarity condition and the monotonicity of the rate-MSE
constraint, the value of $\lambda_{l}^{\star}$ can be computed as
follows: 1) $\lambda_{l}^{\star}=0$ if the rate-MSE constraint is
satisfied with $\lambda_{l}^{\star}=0$; 2) otherwise, perform a
bisection search to obtain the optimal $\lambda_{l}^{\star}$. For
the latter case, the search will terminate when the rate-MSE
constraint is active, i.e., when equality holds true.

\subsubsection{\bf Solving Step 4 for Algorithm 1} The corresponding problem to
update $\{\hat{\bf r},{\bf p}\}$, i.e., step 4 of Algorithm 1, can
be decomposed into two parts. The first part has to do with the flow
rate conservation constraint with optimization variable $\hat {\bf
r}$, and the second part has to do with $\bf{p}$.

 The first part can again be separated into two independent subproblems, one for $
\hat{r}$ and another for the rest of the variables in $\hat{\bf r}$.

{\bf (i) Subproblem for $\hat{r}$}: The subproblem for variable
$\hat{r}$ is given by the following easy unconstraint quadratic
optimization problem
\begin{align}\label{UpdateSecondSub1Sol}
\arg\max ~&\frac{\hat r}{2}-\frac{\rho_{1}}{2}\left(\hat
 r-r-\frac{\delta}{\rho_{1}}\right)^{2}=r+\frac{1+2\delta}{2\rho_{1}}.
\end{align}

{\bf (ii) Subproblem for $\{\hat r_{m}^{S(m)},\hat r_{m}^{D(m)},\hat
r_{l}^{s}(m),\hat r_{l}^{d}(m)\}$}:  In this subproblem, the rest of
the variables in $\hat{\bf r}$ are updated, subject to the
conservation constraints of flow rate. As we have discussed before,
the introduction of the auxiliary local optimization variables, i.e.,
\eqref{SlackLinks}, make this subproblem decoupled over each node
$v\in\mathcal{V}$ and commodity $m$. As such, problem
\eqref{WMMSEQSecond} decomposes into a series of simpler problems,
one for each tuple $(m,l,v)$
\begin{align}\label{UpdateSecondIndividual}
\min~&\sum_{l\in{\rm IN}(v)\cup {\rm
Out}(v)}\left(\hat
r_{l}^{v}(m)-r_{l}(m)-\frac{\delta_{l}^{v}(m)}{\rho_{1}}\right)^{2}+1_{\{S(m),D(m)\}}(v)\left(
 \hat r_{m}^{v}-r_{m}-\frac{\delta_{m}^{v}}{\rho_{1}}\right)^{2}\notag\\
{\rm s. t.}~&\sum_{l\in{\rm In}(v)}\hat
r_{l}^{v}(m)+1_{\{S(m)\}}(v)\hat
r_{m}^{v}=\sum_{l\in{\rm Out}(v)}\hat
r_{l}^{v}(m)+1_{\{D(m)\}}(v)\hat r_{m}^{v}.
\end{align}
Since problem \eqref{UpdateSecondIndividual} has only one equality constraint, it admits a closed-form solution. In particular, let us denote the optimal dual Lagrangian variable as $\lambda_{v}^{\star}(m)$. Using the first-order optimality condition, the optimal solution for \eqref{UpdateSecondIndividual} is given by
\begin{align}\label{UpdateSecondIndividualSol}
\!\!\!&\hat r_{l}^{v\star}(m)=\left\{\begin{array}{cc}
r_{l}(m)+\frac{\delta_{l}^{v}(m)}{\rho_{1}}-\lambda_{v}^{\star}(m),&l\in{\rm Out}(v)\\
r_{l}(m)+\frac{\delta_{l}^{v}(m)}{\rho_{1}}+\lambda_{v}^{\star}(m),& l\in{\rm In}(v)
\end{array}\right.,
\end{align}
and
\begin{align}\label{UpdateSecondIndividualSol2}
\hat r_{m}^{v\star}=\left\{\begin{array}{cc}
r_{m}+\frac{\delta_{m}^{v}}{\rho_{1}}-\lambda_{v}^{\star}(m),& v\in D(m)\\
r_{m}+\frac{\delta_{m}^{v}}{\rho_{1}}+\lambda_{v}^{\star}(m),& v\in S(m)
\end{array}\right.
\end{align}
where
\begin{eqnarray*}
\lambda_{v}^{\star}(m)&=&\Bigg[\sum_{l\in{\rm
Out}(v)}\left(r_{l}(m)+\frac{\delta_{l}^{v}(m)}{\rho_{1}}\right)-\sum_{l\in{\rm
In}(v)}\left(r_{l}(m)+\frac{\delta_{l}^{v}(m)}{\rho_{1}}\right)+\left(r_{m}+\frac{\delta_{m}^{v}}{\rho_{1}}\right) \\
&&\times\left(1_{\{D(m)\}}(v)-1_{\{S(m)\}}(v)\right)\Bigg]\left(|{\rm
In}(v)\cup {\rm Out}(v)|+1_{\{S(m),D(m)\}}(v)\right)^{-1}.
\end{eqnarray*}

 {\bf (iii) Subproblem for ${\bf p}$}: The remaining part is for optimization variable ${\bf p}$ with power budget constraints, and this updating procedure can be decoupled over each BS. For BS $s\in\mathcal{B}$, the updating rule is,
\begin{align}\label{WMMSEQSecondSub}
\min ~&\sum_{k=1}^{K}\sum_{d:l=(s,d,k)\in\mathcal{L}^{wl}\atop n=(s',d',k)\in
\bar I(l)}\left(p_{ds}^{k}-p_{d's',ds}^{k}-\frac{\theta_{nl}}{\rho_{2}}\right)^{2}\notag\\
{\rm s.
t.}~&\sum_{k=1}^{K}\sum_{d:l=(s,d,k)\in\mathcal{L}^{wl}}|p_{ds}^{k}|^{2}\leq
\bar p_{s}.
\end{align}
By denoting the optimal Lagrange dual variable for the power
constraint as $\lambda_{s}^{\star}\geq 0$ and the optimal solution
of problem \eqref{WMMSEQSecondSub} as $\{p_{ds}^{k\star} \mid
(s,d,k)\in\mathcal{L}^{wl}\}$, the first-order optimality condition
can be expressed as
\begin{align}\label{WMMSEQSecondSubSol}
p_{ds}^{k\star}=\frac{\sum_{(s',d',k)\in\bar I(s,d,k)}p_{d's',ds}^{k}+\frac{\theta_{(s',d',k),(s,d,k)}}{\rho_{2}}}{|\bar I(s,d,k)|+\lambda_{s}^{\star}},~k=1\sim K.
\end{align}
Since the following sum
$$\sum_{k=1}^{K}\sum_{d:l=(s,d,k)\in\mathcal{L}^{wl}}\mid p_{ds}^{k\star}|^{2}$$
is a decreasing function of $\lambda_{s}^{\star}$ and the complementarity condition, it follows that $\lambda_{s}^{\star}=0$ if the corresponding constraint is already satisfied $$\sum_{k=1}^{K}\sum_{d:l=(s,d,k)\in\mathcal{L}^{wl}}\mid p_{ds}^{k\star}|^{2}\leq \bar p_{s}.$$
Otherwise, $\lambda_{s}^{\star}$ can be chosen via a bisection search to ensure the power budget constraint is active.

To summarize, all the steps in Algorithm 1 (including the updating
of the Lagrange dual variables, \eqref{UpdateDual}) can
be efficiently computed.

\bibliographystyle{IEEEbib}
\bibliography{refs13}

\end{document}